\documentclass{aa}
\usepackage{natbib}
\bibpunct{(}{)}{;}{a}{}{,} 

\usepackage[english]{babel}           
\usepackage{graphicx,epsfig}
\usepackage{txfonts}
\usepackage{mathrsfs}
\usepackage{lscape}
\usepackage{eucal}
\usepackage{pifont}
\usepackage{rotating}
\usepackage{multirow}
\usepackage{tikz}
\usepackage[colorlinks=true,citecolor=blue,linkcolor=blue]{hyperref}
\usepackage{multicol}
\usepackage{longtable}
\usepackage[fleqn]{amsmath}
\usepackage{breqn}
\usepackage{mathtools}
\usepackage{relsize}

\def \cc    {\ifmmode{\,{\rm cm}^{-3}}\else{$\,{\rm cm}^{-3}$}\fi}
\def \cq    {\ifmmode{\,{\rm cm}^{-2}}\else{$\,{\rm cm}^{-2}$}\fi}
\def \mic   {\ifmmode{\,\mu{\rm m}}\else{$\mu$m}\fi}
\def \eccs  {\ifmmode{\,{\rm erg}~{\rm cm}^{-3}~{\rm s}^{-1}}\else{$\,{\rm erg}~{\rm cm}^{-3}~{\rm s}^{-1}$}\fi}
\def \ecc   {\ifmmode{\,{\rm erg}\,{\rm cm}^{-3}}\else{$\,{\rm erg}\,{\rm cm}^{-3}$}\fi}
\def \ecqs  {\ifmmode{\,{\rm erg}\,{\rm cm}^{-2}\,{\rm s}^{-1}\,{\rm 
             sr}^{-1}}\else{$\,{\rm erg}\,{\rm cm}^{-2}\,{\rm s}^{-1}\,{\rm sr}^{-1}$}\fi}
\def \ecss  {\ifmmode{\,{\rm erg}\,{\rm cm}^{-2}\,{\rm s}^{-1}}\else{$\,{\rm erg}\,{\rm cm}^{-2}\,{\rm s}^{-1}$}\fi}
\def \deg   {\ifmmode{^{\circ}}\else{$^{\circ}$}\fi} 
\def \pc    {\ifmmode{\,{\rm pc}}\else{$\,{\rm pc}$}\fi} 
\def \kms   {\ifmmode{\,{\rm km}\,{\rm s}^{-1}}\else{km~s$^{-1}$}\fi} 
\def \kmspc {\ifmmode{\,{\rm km}\,{\rm s}^{-1}\,{\rm pc}^{-1}}\else{km s$^{-1}$ pc$^{-1}$}\fi} 
\def \MJysr {\ifmmode{\,{\rm MJy\,sr}^{-1}}\else{$\,{\rm MJy\,sr}^{-1}$}\fi} 
\def \Kkms  {\ifmmode{\,{\rm K\,km\,s}^{-1}}\else{$\,{\rm K\,km\,s}^{-1}$}\fi}
\def \epso{\ifmmode{\overline{\varepsilon}_{\rm obs}}\else{$\overline{\varepsilon}_{\rm obs}$}\fi}
\def \utM{\ifmmode{u_{\theta,{\rm M}}}\else{$u_{\theta,{\rm M}}$}\fi}
\def \urM{\ifmmode{u_{r,{\rm M}}}\else{$u_{r,{\rm M}}$}\fi}
\def \twCO{\ifmmode{\rm ^{12}CO}\else{$\rm^{12}CO$}\fi} 
\def \thCO{\ifmmode{\rm ^{13}CO}\else{$\rm^{13}CO$}\fi} 
\def \CeiO{\ifmmode{\rm C^{18}O}\else{$\rm C^{18}O$}\fi} 
\def \twCN{\ifmmode{\rm ^{12}CN}\else{$\rm^{12}CN$}\fi} 
\def \thCN{\ifmmode{\rm ^{13}CN}\else{$\rm^{13}CN$}\fi} 
\def \HdCO{\ifmmode{\rm H_{2}CO}\else{$\rm H_{2}CO$}\fi} 
\def \twHdCO{\ifmmode{\rm ^{12}H_{2}CO}\else{$\rm^{12}H_{2}CO$}\fi} 
\def \thHdCO{\ifmmode{\rm ^{13}H_{2}CO}\else{$\rm^{13}H_{2}CO$}\fi} 
\def \twC{\ifmmode{\rm ^{12}C}\else{$\rm^{12}C$}\fi} 
\def \thC{\ifmmode{\rm ^{13}C}\else{$\rm^{13}C$}\fi} 
\def \Hp{\ifmmode{\rm H^+}\else{$\rm H^+$}\fi} 
\def \Cp{\ifmmode{\rm C^+}\else{$\rm C^+$}\fi} 
\def \Sp{\ifmmode{\rm S^+}\else{$\rm S^+$}\fi} 
\def \Op{\ifmmode{\rm O^+}\else{$\rm O^+$}\fi} 
\def \CFp{\ifmmode{\rm CF^+}\else{$\rm CF^+$}\fi}
\def \CHp{\ifmmode{\rm CH^+}\else{$\rm CH^+$}\fi}
\def \CHdp{\ifmmode{\rm CH_2^+}\else{$\rm CH_2^+$}\fi}
\def \CHtp{\ifmmode{\rm CH_3^+}\else{$\rm CH_3^+$}\fi} 
\def \SHp{\ifmmode{\rm SH^+}\else{$\rm SH^+$}\fi}
\def \SHdp{\ifmmode{\rm SH_2^+}\else{$\rm SH_2^+$}\fi}
\def \SHtp{\ifmmode{\rm SH_3^+}\else{$\rm SH_3^+$}\fi}
\def \twCHp{\ifmmode{\rm ^{12}CH^+}\else{$\rm^{12}CH^+$}\fi}
\def \thCHp{\ifmmode{\rm ^{13}CH^+}\else{$\rm^{13}CH^+$}\fi}
\def \CtH{\ifmmode{\rm C_2H}\else{$\rm C_2H$}\fi} 
\def \CthHt{\ifmmode{\rm C_3H_2}\else{$\rm C_3H_2$}\fi} 
\def \Htp{\ifmmode{\rm H_3^+}\else{$\rm H_3^+$}\fi} 
\def \COp{\ifmmode{\rm CO^+}\else{$\rm CO^+$}\fi} 
\def \HCOp{\ifmmode{\rm HCO^+}\else{$\rm HCO^+$}\fi} 
\def \HtOp{\ifmmode{\rm H_3O^+}\else{$\rm H_3O^+$}\fi} 
\def \HCfiN{\ifmmode{\rm HC_5N}\else{$\rm HC_5N$}\fi} 
\def \wat{\ifmmode{\rm H_2O}\else{$\rm H_2O$}\fi} 
\def \HdO{\ifmmode{\rm H_2O}\else{$\rm H_2O$}\fi} 
\def \OHp{\ifmmode{\rm OH^+}\else{$\rm OH^+$}\fi} 
\def \HdOp{\ifmmode{\rm H_2O^+}\else{$\rm H_2O^+$}\fi} 
\def \HtOp{\ifmmode{\rm H_3O^+}\else{$\rm H_3O^+$}\fi} 
\def \NHd{\ifmmode{\rm NH_2}\else{$\rm NH_2$}\fi} 
\def \NHtrois{\ifmmode{\rm NH_3}\else{$\rm NH_3$}\fi} 
\def \oxy{\ifmmode{\rm O_2}\else{$\rm O_2$}\fi} 
\def \HH{\ifmmode{\rm H_2}\else{$\rm H_2$}\fi}
\def \Jone{\ifmmode{\rm {(J=1--0)}}\else{{(J=1--0)}}\fi} 
\def \Jtwo{\ifmmode{\rm {(J=2--1)}}\else{{(J=2--1)}}\fi} 
\def \Jthr{\ifmmode{\rm {(J=3--2)}}\else{{(J=3--2)}}\fi} 
\def \Jfou{\ifmmode{\rm {(J=4--3)}}\else{{(J=4--3)}}\fi} 
\def \Jfiv{\ifmmode{\rm {J=4--3}}\else{{J=4--3}}\fi} 
\def \Ta{\ifmmode{\rm T_A}\else{$\rm T_A$}\fi} 
\def \Tas{\ifmmode{\rm T_A^*}\else{$\rm T_A^*$}\fi} 
\def \Tmb{\ifmmode{\rm T_{mb}}\else{$\rm T_{mb}$}\fi} 
\def \Tr{\ifmmode{\rm T_r}\else{$\rm T_r$}\fi} 
\def \Trs{\ifmmode{\rm T_r^*}\else{$\rm T_r^*$}\fi}
\def \NHt{\ifmmode{N_{\rm H}}\else{$N_{\rm H}$}\fi}
\def \NH{\ifmmode{N({\rm H})}\else{$N({\rm H})$}\fi}
\def \NH2{\ifmmode{N({\rm H}_2)}\else{$N({\rm H}_2)$}\fi}
\def \NCH{\ifmmode{N({\rm CH})}\else{$N({\rm CH})$}\fi}
\def \NHF{\ifmmode{N({\rm HF})}\else{$N({\rm HF})$}\fi}
\def \dens{\ifmmode{n_{\rm H}}\else{$n_{\rm H}$}\fi}
\def \densini{\ifmmode{n_{\rm H}^0}\else{$n_{\rm H}^0$}\fi}
\def \densfin{\ifmmode{n_{\rm H}^{\rm f}}\else{$n_{\rm H}^{\rm f}$}\fi}
\def \densSNR{\ifmmode{n_{\rm H}^{\rm SN}}\else{$n_{\rm H}^{\rm SN}$}\fi}
\def \nCO{\ifmmode{n({\rm CO})}\else{$n({\rm CO})$}\fi}
\def \nHF{\ifmmode{n({\rm HF})}\else{$n({\rm HF})$}\fi}
\def \nH2{\ifmmode{n({\rm H}_2)}\else{$n({\rm H}_2)$}\fi}

\defcitealias{Jenkins2011}{JT11}
\defcitealias{Godard2024a}{paper~I}

\begin{document}

\title{Shocks in the warm neutral medium}
\subtitle{II - Origin of neutral carbon at high pressure}

\author{
  B. Godard            \inst{\ref{lerma}, \ref{ENS}}, 
  G. Pineau des Forêts \inst{\ref{IAS},   \ref{lerma}}, 
  J. La Porte           \inst{\ref{Sorbonne}}, \and
  M. Merlin-Weck       \inst{\ref{LLG}} 
}

\institute{
Observatoire de Paris, Université PSL, Sorbonne Université, LERMA, 75014 Paris, France
\label{lerma}
\and
Laboratoire de Physique de l’Ecole Normale Supérieure, ENS, Université PSL, CNRS, Sorbonne Université, Université de Paris, F-75005 Paris, France
\label{ENS}
\and
Université Paris-Saclay, CNRS, Institut d’Astrophysique Spatiale, 91405, Orsay, France
\label{IAS}
\and
Sorbonne Université, F-75005 Paris, France
\label{Sorbonne}
\and
Lycée Louis-le-Grand, F-75005 Paris, France
\label{LLG}
}

 \date{Received 25 May 2024 / Accepted 26 June 2024}

\abstract{}{Ultraviolet (UV) lines of neutral carbon observed in absorption in the local diffuse interstellar medium (ISM) have long revealed that a substantial fraction of the mass of the gas lies at a thermal pressure one to three orders of magnitude above that of the bulk of the ISM. In this paper, we propose that this enigmatic component originates from shocks propagating at intermediate ($V_S > 30$ \kms) and high velocities ($V_S \geqslant 100$ \kms) in the Warm Neutral Medium (WNM).}
{Shock waves irradiated by the standard interstellar radiation field (ISRF) are modeled using the Paris-Durham shock code designed to follow the dynamical, thermal, and chemical evolutions of shocks with velocities up to 500 \kms. Each observed line of sight is decomposed into a high pressure and a low pressure components. The column density of carbon at high pressure is confronted to the model predictions to derive the number of shocks along the line of sight and their total dissipation rate.}
{Phase transition shocks spontaneously lead to the presence of high pressure gas in the diffuse ISM and are found to naturally produce neutral carbon with excitation conditions and linewidths in remarkable agreement with the observations. The amounts of neutral carbon at high pressure detected over a sample of 89 lines of sight imply a dissipation rate of mechanical energy with a median of $\sim$~$3 \times 10^{-25}$~\eccs\ and a dispersion of about a factor of three. This distribution of the dissipation rate weakly depends on the detailed characteristics of shocks as long as they propagate at velocities between 30 and 200~\kms\ in a medium with a preshock density $\densini \geqslant 0.3$~\cc\ and a transverse magnetic field $B_0 \leqslant 3$~$\mu$G. We not only show that this solution is consistent with a scenario of shocks driven by supernova remnants (SNR) but also that this scenario is, in fact, unavoidable. Any line of sight in the observational sample is bound to intercept SNRs, mostly distributed in the spiral arms of the Milky Way, and expanding in the diffuse ionized and neutral phases of the Galaxy. Surprisingly, the range of dissipation rate derived here, in events that probably drive turbulence in the WNM, is found to be comparable to the distribution of the kinetic energy transfer rate of the turbulent cascade derived from the observations of CO in the Cold Neutral Medium (CNM).}
{This work reveals a possible direct tracer of the mechanisms by which mechanical energy is injected in the ISM. It also suggests that a still unknown connection exists between the amount of energy dissipated during the injection process in the WNM and that used to feed interstellar turbulence and the turbulent cascade observed in the CNM.}


\authorrunning{B. Godard et al.}
\titlerunning{Phase transition shocks in the warm neutral medium} 
\maketitle

\section{Introduction}

The local interstellar medium is full of chemical mysteries. Since the first detection of molecules in space (e.g. \citealt{Adams1941a,Douglas1941}), observations of atoms and molecules have revealed chemical, excitation, and kinematic properties that regularly challenge idealized descriptions of the diffuse neutral gas. Among these challenges are the observations of \CHp\ (e.g. \citealt{Gredel2002, Pan2004, Sheffer2008}), of \SHp\ \citep{Godard2012}, and of the pure rotational levels of \HH\ (e.g. \citealt{Wakker2006a, Shull2021a}), whose abundances are several orders of magnitude above those predicted by a chemistry solely driven by UV photons, the bimodality of the line profiles of CH \citep{Lambert1990,Crane1995} which suggests that this species is formed through two separated chemical pathways, or the tight correlation between the abundances of HCO$^+$ and OH or H$_2$O \citep{Lucas1996, Gerin2019a} which cannot be explained by static clouds at chemical equilibrium. As shown in several studies, all these observations are precious as they may be diagnostics of the dynamical evolution (condensation and evaporation) of the multiphase  ISM \citep{Valdivia2017, Godard2023}, the processes of dissipation of mechanical energy in the Cold Neutral Medium \citep{Godard2014, Lesaffre2020}, or the out-of-equilibrium chemical evolution of molecular clouds \citep{Panessa2023a}. More generally, these features detected in the local ISM are of fundamental importance to identify the energy sources that power atomic and molecular emission lines which, in turn, can be used to interpret extragalactic observations and draw the mechanical and radiative energy balance of the ISM in entire galaxies \citep{Lehmann2022,Villa-Velez2024a}.

Among these chemical conundrums is the intriguing excitation conditions of neutral carbon observed in the local diffuse interstellar gas. Using the high resolution Space Telescope Imaging Spectrograph aboard the {\it Hubble Space Telescope}, \citet{Jenkins2001a} and \citet{Jenkins2011} (hereafter, \citetalias{Jenkins2011}) conducted several observations of the UV multiplets of CI in absorption against the continuum of nearby OB stars. These absorption lines, which originate from the three fine structure levels of neutral carbon, provide the column averaged ratios of the level populations, hence clues regarding the density and temperature of the foreground absorbing material. Through a detailed analysis of the excitation of neutral carbon, \citetalias{Jenkins2011} showed that these observations can only be explained if the lines of sight are composed of two distinct environments: a medium at low thermal pressure ($P \sim 3.8 \times 10^{3}$ K~cm$^{-3}$), which encompasses $\sim 95$\% of the mass of the gas and corresponds to the bulk of the ISM \citep{Wolfire2003}, and a medium at a thermal pressure one to three orders of magnitude above. The coexistence along every lines of sight of environments with large contrasts in thermal pressure raises a deep interpretative challenge regarding the dynamical state of the diffuse interstellar matter, the origin of the high pressure component, its survival timescale, and the absence of gas at intermediate thermal pressure.

In their analysis, \citetalias{Jenkins2011} originally proposed that the gas at high thermal pressure could originate from the dissipation of turbulent energy in the Cold Neutral Medium (CNM) \citep{Joulain1998, Godard2009}, from the propagation of shock waves in the Warm Neutral Medium (WNM) \citep{Bergin2004a} or, alternatively, from the interactions of the background stars with their surrounding environments which include the expansion of HII regions and the pressure induced by the recoil of H atoms following the photodissociation of \HH\ at the edge of molecular clouds \citep{Field2009a}. In a follow up study, \citet{Jenkins2021a} finally ruled out the influence of the background stars by observing CI in absorption toward extragalactic sources and finding similar fractions of diffuse gas at high thermal pressure along these lines of sight. This lack of influence of the background stars is a key result. It suggests that the excitation of neutral carbon is linked to a more general feature of the dynamical evolution of the diffuse matter.

The fact that shocks associated to Supernovae Remnants (SNR) produce gas at high thermal pressure is well known. Analysis of ultraviolet (UV) absorption lines toward stars in the IC~443 and the Vela SNRs (e.g. \citealt{Ritchey2020a,Ritchey2023a} and reference therein) have long revealed the presence of shocked gas at thermal pressures as high as $\sim 10^6$~K~cm$^{-3}$. A step forward was accomplished by \citet{Bergin2004a} who showed that shocks driven in the diffuse atomic gas by spiral arm density waves, supernovae blast waves, or the infall of material onto the galactic disk could explain the excitation state of neutral carbon observed by \citetalias{Jenkins2011}. Indeed, this work promotes the idea that shocks influence the dynamics of the entire interstellar medium and are detectable toward lines of sight with no obvious associated SNR. If this scenario is confirmed, the observations of CI would provide invaluable information regarding the mechanisms by which mechanical energy is injected in the ISM, which are still uncertain and currently believed to be a mixture of stellar feedback and galactic dynamics including orbital motions and mass accretion onto galactic disks \citep{Brucy2020a}. However, and to our knowledge, the scenario proposed by \citet{Bergin2004a} has never been studied quantitatively.

In this paper, we present a detailed and quantitative analysis of the production and excitation of neutral carbon by shocks propagating at intermediate ($30 < V_S \leqslant 100$~\kms) and high velocities ($V_S > 100$~\kms) in the WNM. These shocks, which incidentally induce phase transition between the WNM and the CNM, are modeled using the most recent version of the Paris-Durham shock code designed to follow the dynamical, thermal, and chemical evolutions of shocks with velocities up to 500 \kms\ (\citealt{Godard2024a}, hereafter \citetalias{Godard2024a}). Following \citetalias{Jenkins2011}, the observations of the absorption lines of CI are decomposed into a low and a high pressure components. The column densities of CI at high pressure is then compared with the predictions of the model to derive the distribution of dissipation rates required to explain the observations.

The observational sample and the decomposition procedure of \citetalias{Jenkins2011} are presented in Sect. \ref{Sect-obs}. The model and its predictions regarding the abundance, the excitation, and the line profiles of CI are presented in Sect. \ref{Sect-mod}. The resulting distribution of dissipation rates and its dependence on the model parameters are shown in Sect. \ref{Sect-grid}. The origin of these shocks which likely, but not exclusively, involves SNRs expanding in the diffuse phases of the Galaxy is discussed in Sect. \ref{Sect-disc} where we highlight the unavoidable nature of SNR surfaces in the observational sample of \citetalias{Jenkins2011}. Some limitations of this work are exposed and discussed in Sect.~\ref{Sect-disc2}. The main conclusions are summarized in Sect.~\ref{Sect-conc}.

\section{Observational sample} \label{Sect-obs}

\subsection{Position and distances} \label{Sect-obs-pos}

\begin{figure}[!h]
\begin{center}
\includegraphics[width=9.0cm,trim = 3.5cm 0.5cm 1.5cm 2.0cm, clip,angle=0]{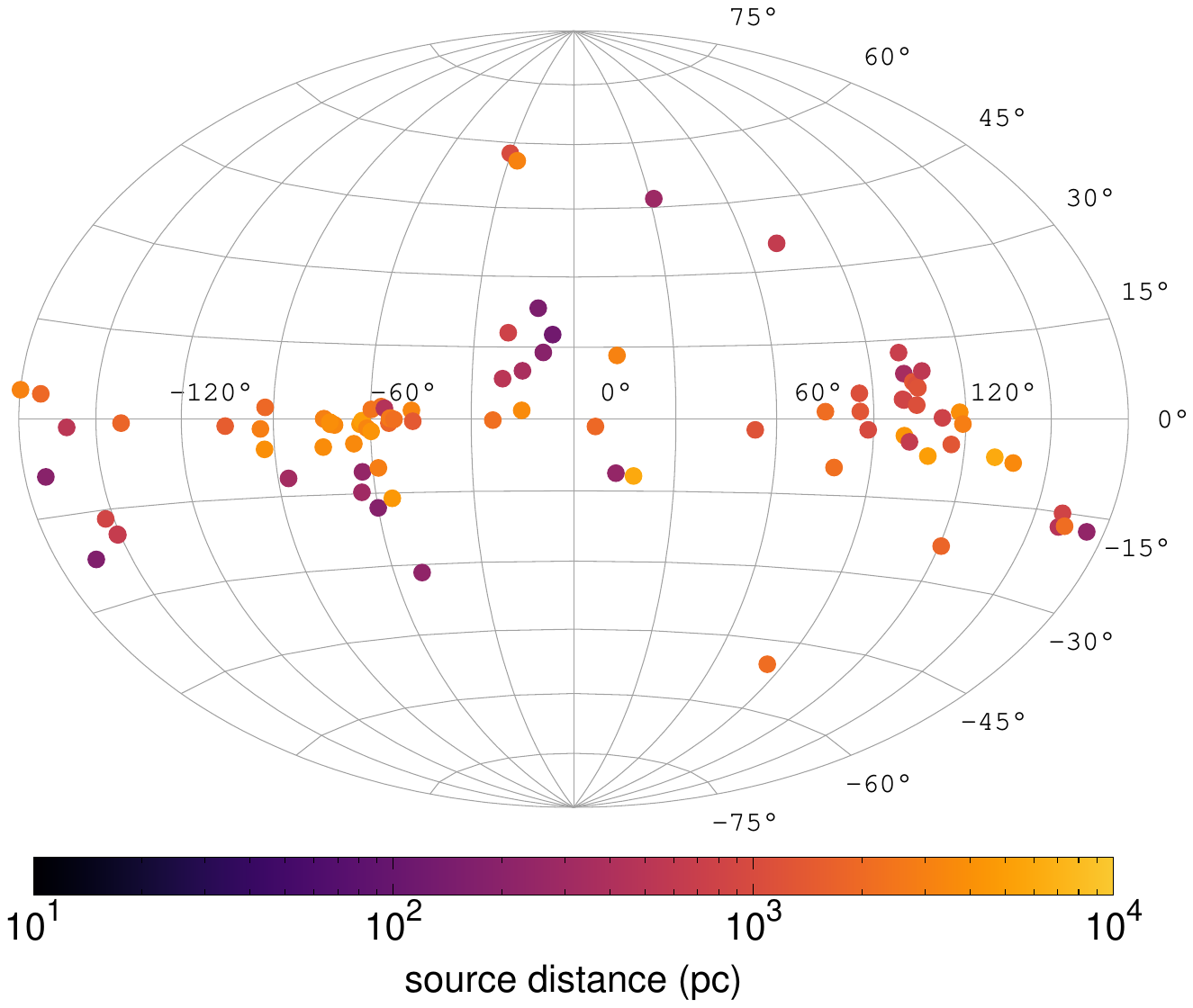}
\caption{Aitoff projection, in Galactic longitude and latitude coordinates, of the background sources of the observational sample of \citetalias{Jenkins2011}. The color-coding indicates the distance of background sources in parsec.}
\label{Fig-Obs-pos}
\end{center}
\end{figure}

The observational dataset studied here contains all the lines of sight observed by \citetalias{Jenkins2011} and available on the Strasbourg astronomical Data Center\footnote{\url{https://cdsarc.cds.unistra.fr/viz-bin/cat/J/ApJ/734/65}}. The sample is composed of 89 lines of sight where the UV multiplets of CI have been detected in absorption against the continuum of nearby OB stars. The position of the sources in Galactic coordinates and their distances derived from measurements of their parallaxes are shown in Fig.~\ref{Fig-Obs-pos}. The background sources cover all Galactic longitudes with two clusters between -120$^{\circ}$ and -30$^{\circ}$ and between 90$^{\circ}$ and 120$^{\circ}$, which are associated to the Scutum-Centaurus, the Sagittarius, and the Perseus spiral arms (see Sect. \ref{Sect-disc}). Similarly to the sample studied by \citet{Bellomi2020}, most of the background sources are located at Galactic latitudes below 15$^{\circ}$, as dictated by the distribution of stars in the solar neighborhood.

The background stars are located at distances between 120~pc and 6.1~kpc. Since the amount of gas exponentially decreases as a function of the distance from the midplane of the Galaxy \citep{Dickey1990a}, the length, $l_{\rm los}$, of the neutral material intercepted by any line of sight is necessarily smaller than the distance of the background source, especially when the source is located at high Galactic latitude. Following the simple prescription adopted by \citet{Bellomi2020} and \citet{Godard2023}, we assume that the gas extends above the midplane over a height of 100 pc and estimate this length as
\begin{equation} \label{Eq-length}
l_{\rm los} = {\rm min}\left(d, \frac{100}{{\rm sin}(|b|)} \right)\,\, {\rm pc},
\end{equation}
where $d$ is the distance of the background source and $b$ is its Galactic latitude. With this prescription, the length of the intercepted material, $l_{\rm los}$, is found to follow a flat distribution in logscale between about 200 pc and 3 kpc. This length will be used in Sects. \ref{Sect-grid} and \ref{Sect-disc} to compute the distribution of dissipation rate in shocks propagating in the WNM. The impact of the prescription chosen to compute $l_{\rm los}$ will be discussed then.

\subsection{Excitation conditions deduced from CI lines}

\begin{figure}[!h]
\begin{center}
\includegraphics[width=9.0cm,trim = 2cm 2.5cm 0.5cm 0.8cm, clip,angle=0]{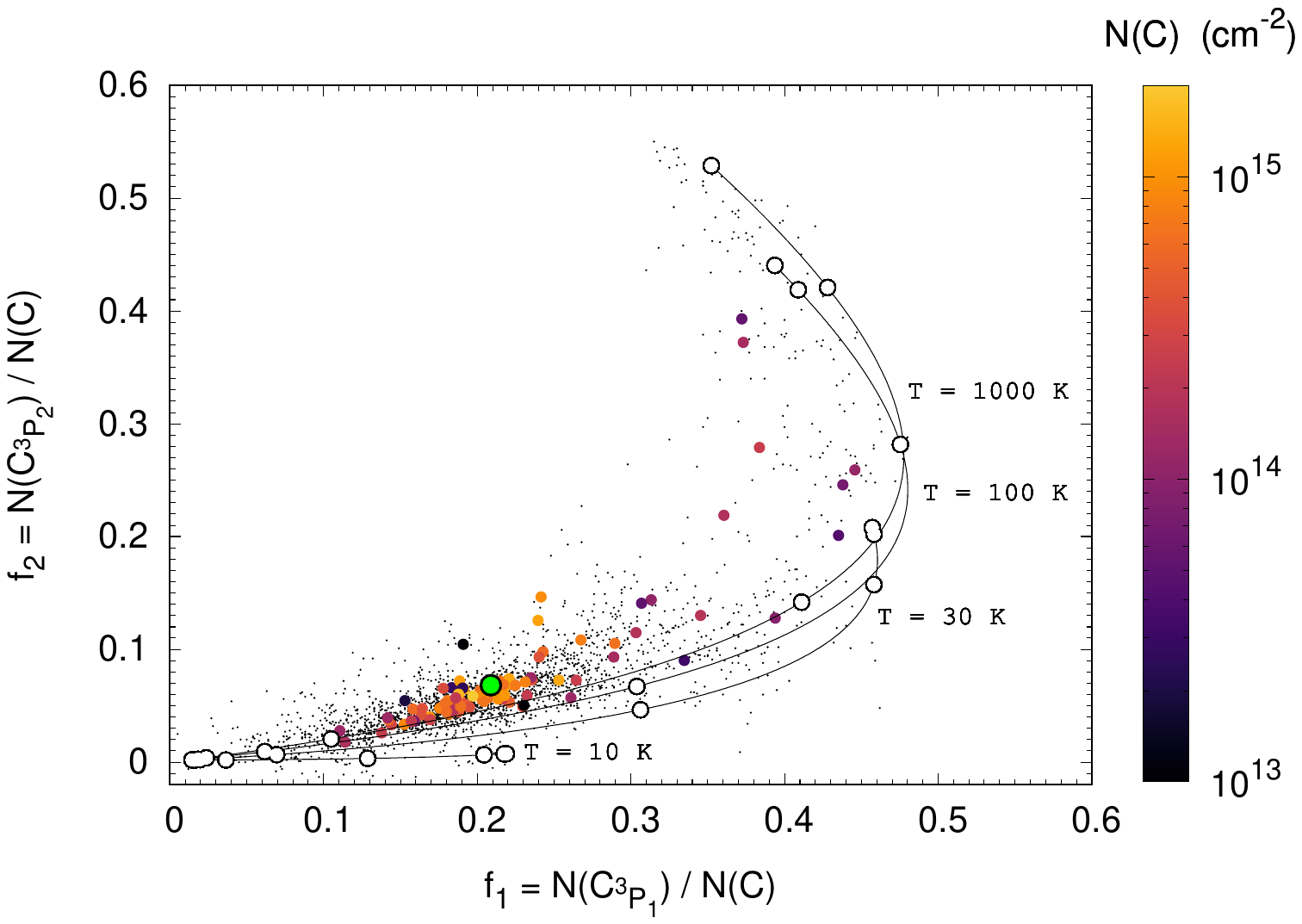}
\caption{Excitation properties of the neutral carbon along the lines of sight observed by \citetalias{Jenkins2011}. The filled colored circles indicate the $f_1$ and $f_2$ column density ratios observed along each line of sight. Each circle is color-coded according to the total column density of neutral carbon, $N({\rm C})$. The black dots show the $f_1$ and $f_2$ column density ratios obtained per velocity interval of 0.5 \kms. The filled green circle shows the total $f_1$ and $f_2$ ratios computed over the entire observational sample. The black curves display the theoretical ratios obtained for homogeneous gas with kinetic temperatures of 10, 30, 100, and 1000~K and thermal pressure varying, along each curve, between $10^2$ and $10^7$~K~\cc (from the bottom left to the top right). The empty white circles indicate integer values of the logarithm of the thermal pressure (see Fig. \ref{Fig-Decomp}).}
\label{Fig-Jenkins}
\end{center}
\end{figure}

The UV spectra observed by \citetalias{Jenkins2011} are composed of absorption lines originating from the three fine structure levels of the ground electronic state of neutral carbon, the $^3P_0$ fundamental level, and the $^3P_1$ and $^3P_2$ excited levels lying respectively at 23.62~K and 62.46~K above the fundamental level. Each spectrum contains several UV multiplets which can be used to solve the confusion induced by the mixture of line opacities. Using this information and a method devised by \citet{Jenkins2001a}, \citetalias{Jenkins2011} derived the column density spectra of each level, per velocity interval of 0.5 \kms, along any line of sight. These final products of their analysis (see Tables 3 and 4 in their paper) are the data used in this work. Following partially their notation, we define $N({\rm C}_{^3P_0})$, $N({\rm C}_{^3P_1})$, and $N({\rm C}_{^3P_2})$ as the column densities of the three fine structure level of carbon integrated over the entire line of sight, and 
\begin{equation}
f_1 = \frac{N({\rm C}_{^3P_1})}{N({\rm C})}
\end{equation}
and
\begin{equation}
f_2 = \frac{N({\rm C}_{^3P_2})}{N({\rm C})}
\end{equation}
as the ratios of the column densities of the two excited levels to the total column density of carbon
\begin{equation}
N({\rm C}) = N({\rm C}_{^3P_0}) + N({\rm C}_{^3P_1}) + N({\rm C}_{^3P_2}).
\end{equation}

Fig.~\ref{Fig-Jenkins} displays the $f_1$ and $f_2$ ratios computed along all lines of sight of the observational sample (colored points), the ratios computed per velocity interval (small black points), and the total ratios obtained over the entire sample (green point). For comparison, Fig.~\ref{Fig-Jenkins} also shows the theoretical $f_1$ and $f_2$ ratios obtained in homogeneous clouds at different proton densities, \dens, and kinetic temperatures, $T$, calculated at equilibrium taking into account excitation and deexcitation of neutral carbon by radiative decay, collisions with electrons, H, \HH, and He, radiative pumping of the electronic states of C, and chemical pumping of the fine structure levels of C during the recombination of \Cp\ (see Sect.~\ref{Sect-mod} for more details). This convoluted figure is similar to Fig.~2 of \citetalias{Jenkins2011}. Although the exact predictions of the theoretical model depends on the abundances of the collisional partners and on the strength of the impinging UV radiation field\footnote{The theoretical predictions displayed in Fig.~\ref{Fig-Jenkins} are obtained assuming $n(\HH)/\dens = 0.5$, $n({\rm He})/\dens = 0.1$, and $n({\rm e}^-)/\dens = 10^{-4}$, and that each point in the homogeneous cloud is irradiated by the UV radiation field of Mathis \citep{Mathis1983} which sets the strengths of the radiative and chemical pumpings.}, the results obtained here are generic and reveal three main features.

First, Fig.~\ref{Fig-Jenkins} shows that almost none of the observed excitation conditions can be explained with an homogeneous cloud at a single thermal pressure. While it is not explicitly shown in the figure, this result holds regardless of the abundances of the collisional partners and of the strength of the UV radiation field (see \citetalias{Jenkins2011}). Second, lines of sight with larger column densities of neutral carbon display $f_1$ and $f_2$ ratios that lie closer to the total ratios derived over the entire sample (green point). This feature is indicative that all lines of sight intercept similar environments with excitation properties that tend toward the mean values when the column density increases by virtue of the central limit theorem. Last, the theoretical predictions of the $f_1$ and $f_2$ ratios obtained at different kinetic temperature and thermal pressure (from $10^2$ to $10^7$~K~\cc) appear to form an enveloppe around the observational points. All these features imply that any line of sight is necessarily composed of gas with markedly different physical conditions and that the observed column densities result from the sum of these different environments.

\subsection{Decomposition method of Jenkins \& Tripp} \label{Sect-decomp-method}

\begin{figure}[!h]
\begin{center}
\includegraphics[width=9.0cm,trim = 2cm 5.1cm 4.0cm 2.0cm, clip,angle=0]{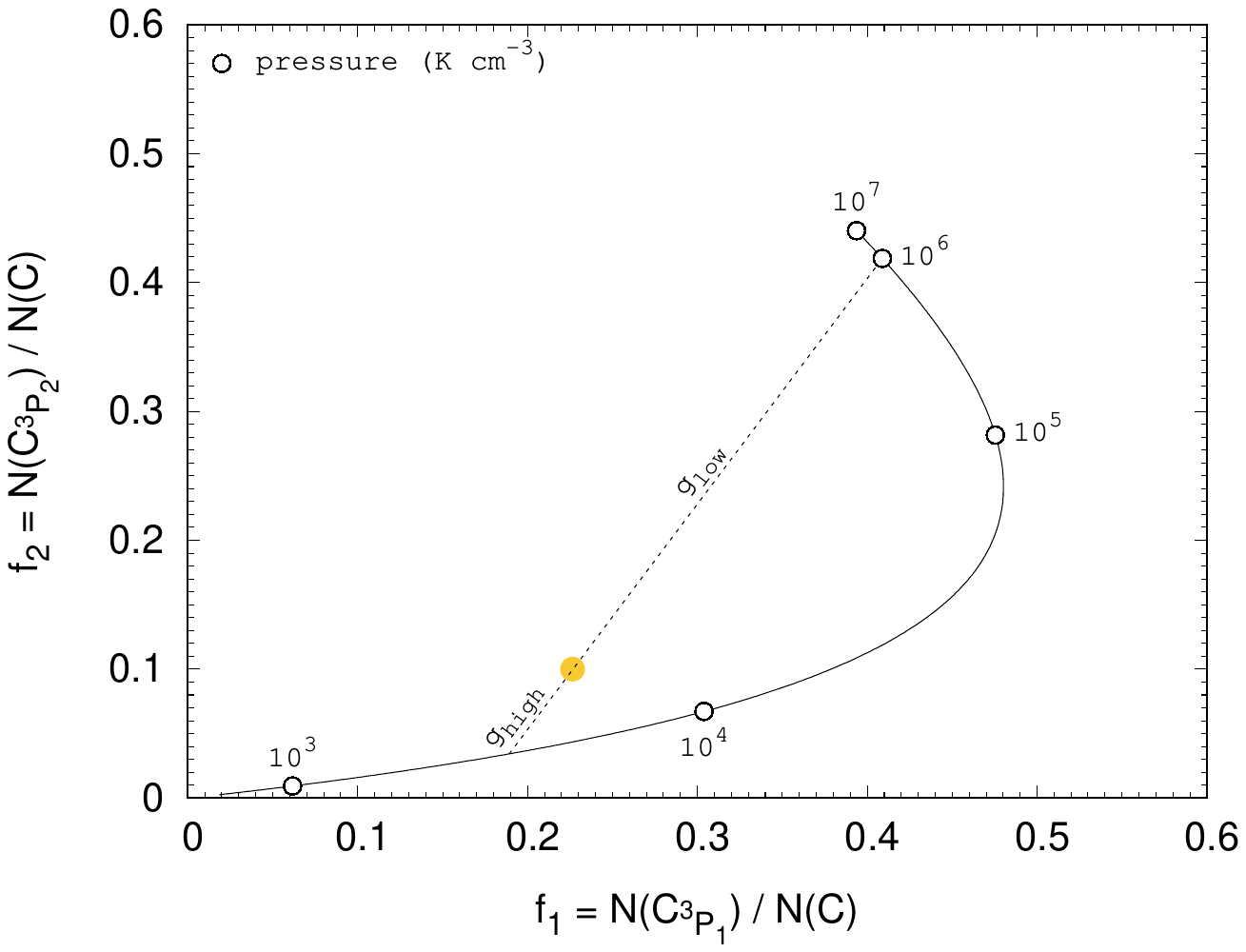}
\includegraphics[width=9.0cm,trim = 2cm 2.8cm 4.0cm 2.0cm, clip,angle=0]{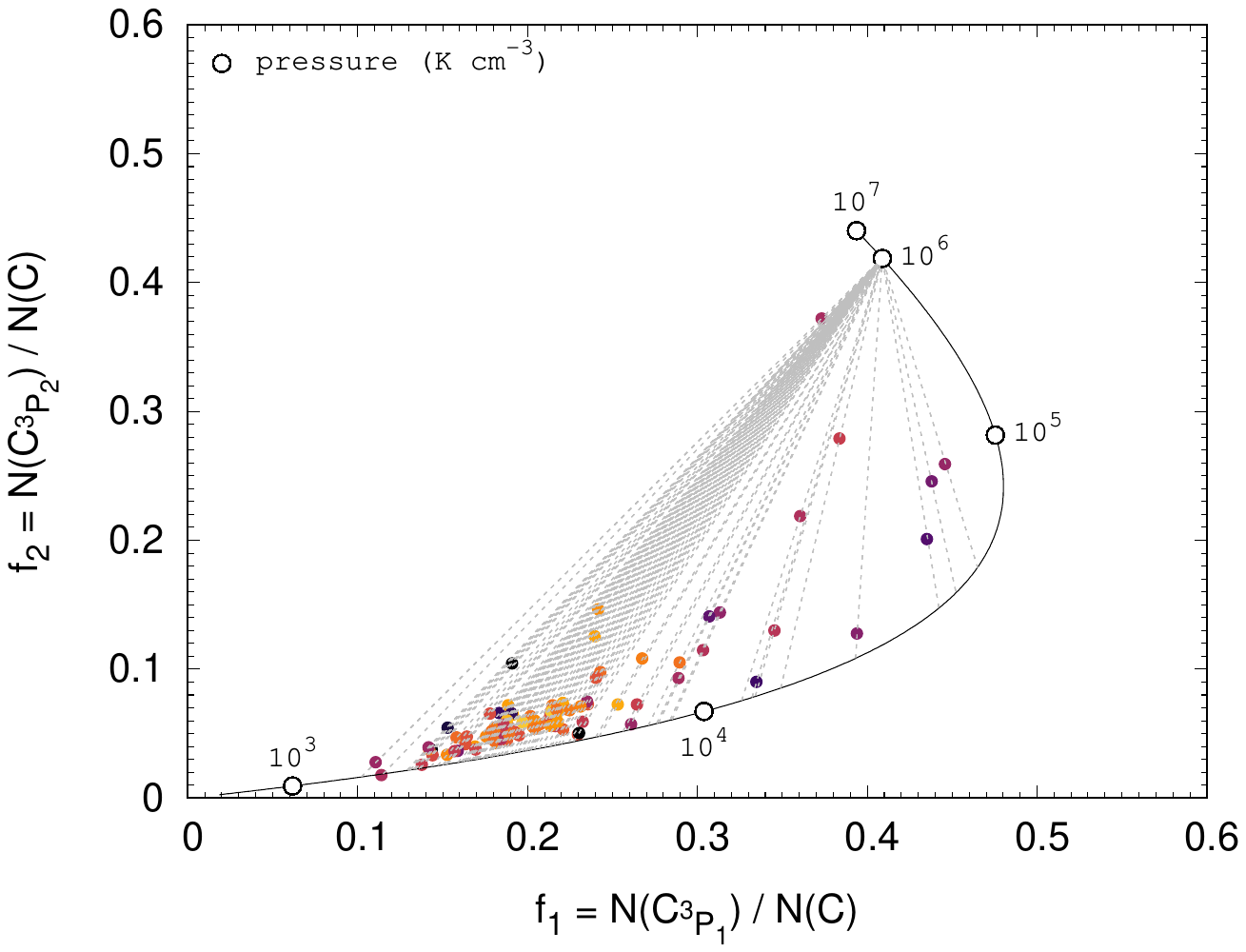}
\caption{Schematic view of the decomposition method proposed by \citetalias{Jenkins2011}. The $f_1$ and $f_2$ column density ratios observed along a given line of sight (top panel) are supposed to result from the linear combination of a fraction $g_{\rm low}$ of neutral carbon at low pressure, and a fraction $g_{\rm high}$ of neutral carbon at high pressure. The thermal pressure of the high pressure components is supposed to be known ($10^6$~K~\cc\ in this example). $g_{\rm low}$ and $g_{\rm high}$ are derived from the geometric position of the observed point along the line that connects the low pressure and high pressure components (dashed black line in the top panel). This decomposition procedure is applied to all observations (bottom panel) to derive the distribution of thermal pressure of the low pressure component.}
\label{Fig-Decomp}
\end{center}
\end{figure}

Through a detailed analysis, \citetalias{Jenkins2011} demonstrated (see Fig. 3 of their paper) that the excitation conditions displayed in Fig. \ref{Fig-Jenkins} cannot result from a lognormal distribution of thermal pressure expected in a turbulent environment (e.g. \citealt{Kritsuk2007}) or from a power-law tail of thermal pressure originating from polytropic gas with an exponent smaller than 1 \citep{Passot1998a} or self-gravitating or collapsing clouds \citep{Federrath2013a,Girichidis2014a}. The reason is that the curvature of the line defined by the $f_1$ and $f_2$ combinations as function of the gas thermal pressure (see Fig.~\ref{Fig-Jenkins}) is insufficient to raise the $f_2$ ratio to the observed values if a lognormal or a power-law distributions of pressure are assumed. They concluded that the most straightforward interpretation of the data is to consider that any line of sight results from a bimodal distribution where most of the gas lies at thermal pressures between $\sim 10^3$ and $10^4$~K~\cc\ and a small fraction of the mass at thermal pressures one to three order of magnitude above.

Expanding on this conclusion, \citetalias{Jenkins2011} applied the decomposition procedure schematized in Fig.~\ref{Fig-Decomp}. They assumed that the $f_1$ and $f_2$ ratios of the high pressure gas are known (and correspond, for instance, to the values derived for $T = 1000$~K and $P = 10^6$~K~\cc, see Fig. \ref{Fig-Decomp}). If the kinetic temperature of the low pressure environment is known (the value used in the theoretical model is fixed in Fig.~\ref{Fig-Decomp} for pedagogical purposes), any observed line of sight (yellow point on the top panel of Fig. \ref{Fig-Decomp}) can be seen as  as a linear combination of a fraction $g_{\rm low}$ of neutral carbon at low pressure and a fraction $g_{\rm high}$ of neutral carbon at high pressure. \citetalias{Jenkins2011} estimated the kinetic temperature of the gas at low pressure from measurements of the rotational temperature of the two first levels of molecular hydrogen (see Table 2 in their paper). When this temperature was not measured, they assumed a canonic value of 80~K close to the mean kinetic temperature of the CNM. Applying this methodology to the entire sample (bottom panel of Fig. \ref{Fig-Decomp}), \citetalias{Jenkins2011} derived the $g_{\rm low}$ and $g_{\rm high}$ fractions along every lines of sight and the distribution of thermal pressure of the low pressure environment. By doing so, they not only found that $\sim 95$\% of the mass of the gas lies at low pressure with a lognormal distribution centered at $P \sim 3.8 \times 10^3$~K~\cc, but they also demonstrated that these findings weakly depends on the exact $f_1$ and $f_2$ ratios assumed for the high pressure component.

Despite these successes, the observations of the excited levels of neutral carbon still raise several conundrums. Although the exact $f_1$ and $f_2$ ratios of the high pressure component weakly affects the decomposition, it would be interesting to know what are the actual physical conditions of this gas. More importantly, these observations raise the question of the origin of the high pressure component. Why should the interstellar medium contains very small amounts of gas ($\sim 5$\%) at anomalously large pressures and how is it possible that such component, seen on every lines of sight, coexists with CNM environments at much lower thermal pressure ?

\subsection{Line profiles}

\begin{figure}[!h]
\begin{center}
\includegraphics[width=9.0cm,trim = 2cm 2.8cm 2.0cm 0.8cm, clip,angle=0]{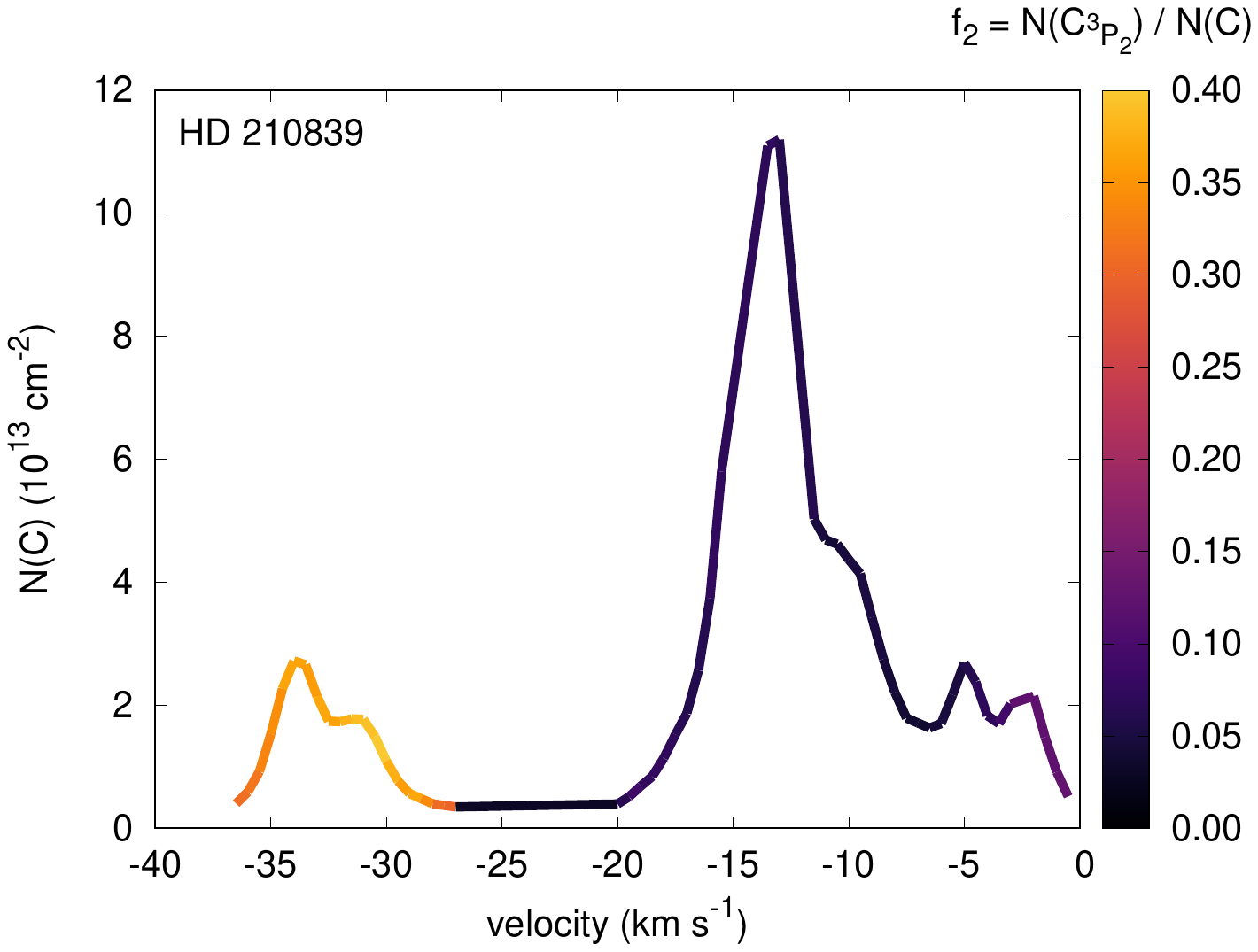}
\caption{Column density spectrum of neutral carbon observed toward HD~210839 (from Table 4 of \citetalias{Jenkins2011}). The column densities are derived over velocity intervals of 0.5 \kms. Each point along the spectrum is color-coded according to the value of the $f_2$ column density ratio to highlight velocity components at different thermal pressures.}
\label{Fig-HD210839}
\end{center}
\end{figure}

\begin{table}
\begin{center}
\caption{List and properties of unblended velocity components observed at high pressure in the sample of \citetalias{Jenkins2011}. These are identified as isolated velocity components where the $f_2$ column density ratio is larger than 0.13 over the entire component (see Fig. \ref{Fig-HD210839} for instance). Numbers in parenthesis are power of 10.}
\label{Tab-obs-prof}
\begin{tabular}{l r r c c c c}
\hline
\multicolumn{2}{c}{source} & velocity & FWHM & N(C)      & $f_1$ & $f_2$\\
                 &         & \kms     & \kms & cm$^{-2}$ \\
\hline
HD  &  37021 &  18.5 &  4.0 & 2.4 (13) & 0.38 & 0.42 \\
    &        &  22.4 &  2.5 & 7.4 (12) & 0.34 & 0.26 \\
HD  &  37061 &  22.7 &  3.1 & 4.7 (13) & 0.36 & 0.40 \\
    &        &  25.9 &  3.3 & 3.2 (13) & 0.40 & 0.32 \\
HD  &  37903 &  27.7 &  3.8 & 1.4 (14) & 0.40 & 0.29 \\
HD  &  71634 &  15.9 &  5.3 & 1.2 (14) & 0.34 & 0.13 \\
HD  &  93205 & -83.2 &  3.3 & 2.6 (13) & 0.35 & 0.49 \\
HD  &  93222 & -25.5 &  5.8 & 2.3 (13) & 0.42 & 0.43 \\
    &        & -21.4 &  3.5 & 1.4 (13) & 0.43 & 0.36 \\
HD  &  93843 & -29.0 &  2.4 & 3.2 (13) & 0.42 & 0.23 \\
HD  & 106343 &  15.6 &  2.1 & 1.9 (13) & 0.39 & 0.18 \\
HD  & 112999 &  -2.6 &  3.3 & 1.9 (13) & 0.42 & 0.24 \\
HD  & 140037 &   9.2 &  3.1 & 6.8 (13) & 0.44 & 0.24 \\
HD  & 147888 &  -8.4 &  5.0 & 2.5 (14) & 0.38 & 0.23 \\
HD  & 148594 &  -4.8 &  4.8 & 1.2 (14) & 0.45 & 0.26 \\
HD  & 210839 & -34.0 &  2.6 & 7.1 (13) & 0.42 & 0.35 \\
    &        & -31.0 &  2.9 & 5.0 (13) & 0.40 & 0.38 \\
HD  & 303308 & -36.4 &  2.8 & 9.3 (12) & 0.39 & 0.48 \\
    &        & -32.8 &  2.2 & 5.5 (12) & 0.37 & 0.48 \\
    &        & -16.9 &  4.3 & 2.3 (13) & 0.33 & 0.52\\
\hline
\end{tabular}
\end{center}
\end{table}

The observed spectra also contains an important piece of information: the velocity structure of the high pressure component. Using the highest resolution configuration of the STIS instrument aboard the {\it Herschel Space Telescope}, \citetalias{Jenkins2011} derived the column density spectra of each level with a resolution of 0.5~\kms. Since the high pressure component is characterized by large values of the $f_2$ ratio, these spectra can be used to put constraints on the kinematic structure of the associated gas. Analysis of the excitation conditions obtained in each channel of the observed spectra shows that, in most of the cases, the velocity components of the gas at high pressure are blended with those of the gas at low pressure. They are, however, a few exception.

Fig.~\ref{Fig-HD210839} displays the column density spectrum of neutral carbon obtained toward HD~210839. While the velocity components above -25~\kms\ are found to have $f_2$ ratios below 0.1 and are thus mostly associated to the low pressure component, the two velocity components below -25~\kms\ are characterized by $f_2$ ratios larger than 0.3 and are therefore entirely composed of gas at high thermal pressure. Expanding on this example, we find 14 lines of sight where the high pressure gas appears as a separated velocity component. Those are identified as isolated velocity components where the $f_2$ column density ratio is larger than 0.13 over the entire component, that is, for each velocity channel. The kinematic information are extracted by performing a multi-Gaussian fits of these 14 spectra using the curve fit optimization algorithm provided by the SciPy python library. The names of these 14 lines of sight, the centroid velocities, the full width at half maximum (FWHM), and the column density of neutral carbon associated to the velocity components at high pressure are given in Table~\ref{Tab-obs-prof}. This table shows that the gas at high pressure display Gaussian velocity profiles with a mean FWHM of 3.5 \kms\ and a dispersion of 1.1 \kms. Interestingly the column density of carbon contained in each velocity profile covers a wide range of values from $5.5 \times 10^{12}$ to a $2.5 \times 10^{14}$ cm$^{-2}$. 

In this paper, we explore the possibility that the high pressure component originates from shock propagating in the WNM. The question is whether interstellar shocks propagating in neutral and diffuse gas can produce neutral carbon in sufficient quantity, with excitation and kinematic properties that match those derived from the observations.

\section{Models of interstellar shocks} \label{Sect-mod}

\subsection{Physical ingredients}

\begin{table*}
\begin{center}
\caption{Main parameters of the shock code, standard model, and range of values explored in this work.}
\label{Tab-main}
\begin{tabular}{l r r r l}
\hline
name & standard & range & unit & definition \\
\hline
$n_{\rm H}^0$  & 1.0                 & $10^{-1}$ $-$ $2$ & cm$^{-3}$           & pre-shock proton density$^a$ \\
$G_0$          & 1                   & $-$               &                     & radiation field scaling factor$^b$ \\
$\zeta_{\HH}$  & $3 \times 10^{-16}$ & $-$               & s$^{-1}$            & \HH\ cosmic ray ionization rate \\
$V_{\rm s}$    & 80                  & 10 $-$ 200        & km s$^{-1}$         & shock velocity \\
$B_0$          & 1.0                 & 0.1 $-$ 10        & $\mu$G              & initial transverse magnetic field \\
\hline
\end{tabular}
\begin{list}{}{}
(a) defined as $n_{\rm H}^0=n^0({\rm H}) + 2n^0(\HH) + n^0(\Hp)$, where $n^0({\rm H})$, $n^0(\HH)$, and $n^0(\Hp)$ are the initial densities of H, \HH, and \Hp.\\
(b) the scaling factor $G_0$ is applied to the standard ultraviolet radiation field of \citet{Mathis1983}.
\end{list}
\end{center}
\end{table*}

Interstellar shocks propagating in the WNM are modeled using the Paris-Durham shock code, a public multi-fluid model\footnote{Available on the ISM plateform \url{https://ism.obspm.fr}} built to follow the dynamical, thermal and chemical structures of shock waves, at steady-state, in a plane-parallel geometry \citep{Godard2019}. Designed to study low velocity shocks ($V_S \leqslant 30$~\kms) propagating in irradiated environments, the code was recently improved by \citet{Godard2024a} (\citetalias{Godard2024a}) to treat shocks propagating at higher velocities, up to 500~\kms, which are known to generate UV, EUV and X-ray photons that interact with the preshock medium and induce the formation of a radiative precursor (e.g. \citealt{Raymond1979, Hollenbach1989, Sutherland2017}). This latest version of the code includes the chemical evolution of multi-ionized species, the cooling induced by these species\footnote{As described in \citetalias{Godard2024a}, the cooling induced by multi-ionized species is computed using the energy levels, radiative transition probabilities, and collisional excitation rates of the CHIANTI database \citep{Dere1997,Dere2019,Del-Zanna2021} available at \url{https://www.chiantidatabase.org}.}, and an exact radiative transfer algorithm for line emission, built on the coupled escape probability formalism \citep{Elitzur2006a}, that accurately follows the propagation and interactions of the photons produced by the shock with the shocked and the preshocked gas.

The Paris-Durham shock code models a plane-parallel shock propagating at a velocity $V_S$ in a magnetized medium with a constant proton density $\densini$ and a magnetic field $B_0$ set in the direction perpendicular to the direction of propagation. The preshock medium and the shocked gas are assumed to be irradiated by an isotropic UV radiation field set to the standard interstellar radiation field of \citet{Mathis1983} and scaled with a parameter $G_0$, and to be pervaded by cosmic ray particles with an equivalent \HH\ ionization rate $\zeta_{\HH}$. Within this geometry and in the frame of reference of the shock front, the code solves the out-of-equilibrium dynamical, thermal, and chemical evolution of a fluid particle during its trajectory from the preshock to the postshock.

The chemical evolution of the abundance of neutral carbon results from a chemical network involving 209 species and 3340 chemical reactions (see \citetalias{Godard2024a}), which include, in particular, photoionization processes \citep{Heays2017}, radiative and dielectronic recombinations \citep{Badnell2003,Badnell2006}, and recombinations onto grains and PAHs \citep{Draine1987}. The populations of the three fine structure levels of C are computed, at each point of the trajectory, taking into account excitation and deexcitation processes by radiative decay, inelastic collisions, radiative pumping and chemical pumping. Inelastic collisions are treated including the collisional rates of neutral carbon with H \citep{Launay1977a}, \HH\ in its ortho and para states \citep{Schroder1991a}, He \citep{Staemmler1991a}, H$^+$ \citep{Roueff1990a}, and e$^-$ \citep{Johnson1987a}. Excitation rates of the fine structure levels by radiative pumping of the electronic states of carbon followed by fluorescence are computed using the optical pumping rates of \citetalias{Jenkins2011} (see Table~6 in their paper). The probability of exciting the neutral carbon during its chemical formation is finally calculated assuming that the three fine structure levels are equally likely to be populated during the recombination of C$^+$, the main formation pathway of C in the models explored here.

\subsection{Standard setup and grid of models}

The range of parameters explored in this work and the values adopted in the standard model are given in Table \ref{Tab-main}. As a prototypical case, we consider a shock propagating at a velocity $V_S=80$~\kms\ in a medium with physical conditions close to those of the WNM observed in the local ISM. The preshock gas is assumed to have a proton density $\densini=1$~\cc, a transverse magnetic field strength $B_0=1$~$\mu$G \citep{Frick2001a,Crutcher2010}, and to be irradiated by the standard UV radiation field of \citet{Mathis1983} ($G_0=1$). The total cosmic ray ionization rate of \HH\ is set to $3\times 10^{-16}$~s$^{-1}$ (e.g. \citealt{Indriolo2015,Neufeld2017}). To explore the impact of various types of shocks on the production and excitation of neutral carbon, we also consider a grid of 325 models, with an initial preshock density varying between 0.1 and 2~\cc, a transverse magnetic field varying between 0.1 and 10 $\mu$G, and a shock velocity ranging between 10 and 200~\kms. The grid of models explored here is similar to the grid presented in \citetalias{Godard2024a}, except for the shock velocity which is limited at 200~\kms. The reason is that the dynamical timescales of SNRs with terminal velocities above 200~\kms\ are shorter than the shock cooling times (see Sect.~\ref{Sect-disc} for more details). This implies that shocks above 200~\kms\ driven by SNRs should not be considered at steady-state.

\begin{figure}[!h]
\begin{center}
\includegraphics[width=9.0cm,trim = 1.5cm 2.2cm 0.5cm 2.2cm, clip,angle=0]{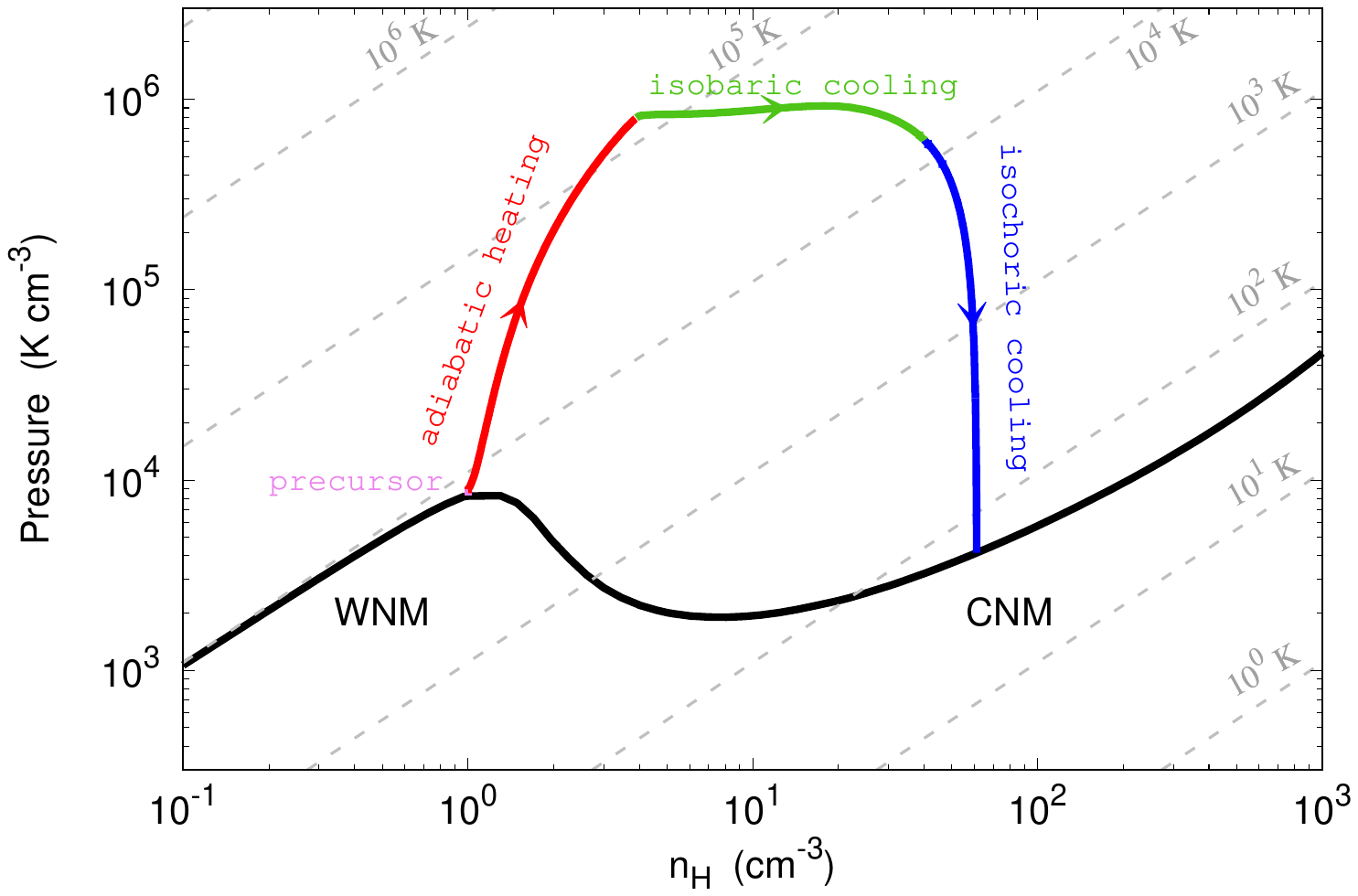}
\caption{Trajectory of a shocked fluid particle from the ambient medium to the postshock region obtained in the standard model. The trajectory is displayed in a proton density $-$ thermal pressure diagram to highlight the phase transition from the WNM to the CNM induced by a shock at 80~\kms. A fluid particle initially evolves through the radiative precursor (pink). As it crosses a shock front, it undergoes three successive regimes: an adiabatic heating (red curve) followed by a quasi-isobaric cooling (green curve), and, finally, a quasi-isochoric cooling (blue curve) as the magnetic pressure becomes dominant in the postshock gas. The black curve indicates the thermal equilibrium state of the diffuse gas obtained for $G_0=1$ (see Fig.~1 of \citetalias{Godard2024a}). Light grey lines are isothermal contours from 1 to $10^6$ K (from bottom right to top left). Note that these isocontours are not evenly spaced due to the change of the number of particles in a gas at high temperature (see Fig.~5 of \citetalias{Godard2024a}).}
\label{Fig-Trajectory-Main}
\end{center}
\end{figure}

The thermodynamical structure of the standard model is summarized in Fig.~\ref{Fig-Trajectory-Main} which displays the trajectory of a shocked fluid particle in a proton density $-$ thermal pressure diagram. As described in \citetalias{Godard2024a}, a fluid particle crossing a shock front undergoes three successive regimes. For strong magnetohydrodynamic shocks, that is shocks where the sonic and the Alfvénic Mach numbers are much higher than one, the fluid particle is first heated adiabatically (red curve in Fig.~\ref{Fig-Trajectory-Main}) to a thermal pressure that scales as
\begin{equation} \label{Eq-postpress}
P = 8.1 \times 10^{5}\ {\rm K}\ \cc\ \left(\frac{\densini}{1\ \cc}\right)\ \left(\frac{V_S}{80\ \kms}\right)^2.
\end{equation}
To ensure an equilibrium between the postshock thermal pressure and the preshock ram pressure, the gas then cools down isobarically (green curve in Fig.~\ref{Fig-Trajectory-Main}). Because the magnetic field is frozen in the ionized fluid, the strength of the magnetic field increases. When the magnetic pressure becomes comparable to the postshock thermal pressure, the gas progressively shifts from an isobaric to an isochoric cooling (blue curve in Fig.~\ref{Fig-Trajectory-Main}). The final density of the gas, $\densfin$, is therefore given by the equilibrium between the postshock magnetic pressure and the preshock ram pressure and writes
\begin{equation} \label{Eq-densfin}
\densfin = 61\ \cc\ \left(\frac{\densini}{1\ \cc}\right)^{3/2} \left(\frac{V_S}{80\ \kms}\right)\ \left(\frac{B_0}{1\ \mu{\rm G}}\right)^{-1}.
\end{equation}
This simple description shows that interstellar shocks with velocities larger than 30~\kms\ bring the gas to thermal pressure larger than $10^5$~K~\cc, and that the gas at high pressure reaches densities comparable to the typical densities of the CNM. Both features are of paramount importance for the production and the excitation of neutral carbon.

It should be noted that this description only holds for strong J-type shocks and cannot be blindly applied to C-type and CJ-type shocks, or even weak J-type shocks where the built up of magnetic pressure during the adiabatic jump suppresses the isobaric cooling stage (see Fig.~8 of \citetalias{Godard2024a}). Keeping this warning in mind, Eq.~\ref{Eq-postpress} is found to accurately describe a large majority of the models explored here, while Eq.~\ref{Eq-densfin} holds for almost all models.

\subsection{Shock size and timescale}

An important issue for the comparison with observations is the calculation of the extent of the shock. By construction, one dimensional stationary models postulate that the gas is confined and cannot escape in the directions parallel to the shock front. As a result, all the shocked gas cools down homogeneously toward thermal equilibrium and remains on this state indefinitely. Such a prescription raises two problems. First, it prevents the development of thermal and dynamical instabilities which are known to occur in the wake of shocks propagating in multiphase environments (e.g. \citealt{Falle2020a, Raymond2020a, Kupilas2021a, Markwick2021a}). Second, it implies that all integrated quantities, such as column densities or line intensities, depend on the extent of the postshock gas (e.g. \citealt{Gusdorf2008a, Leurini2014}).

Since the idealized trajectory computed by the model becomes, at some point, unrealistic, it is necessary to apply a criterion to cut the contribution of the infinite postshock material. Several criteria have been proposed in the past, based on the abundances and temperature profiles or on energetic considerations (e.g. \citealt{Wardle1999, Melnick2015, Godard2019}). In this work, we propose a new criterion based on the physical evolution of the gas. As described in the previous section, shocks propagating in the WNM produce a postshock medium which eventually becomes dominated by magnetic pressure. When it happens, the gas effectively hit a magnetic wall and looses the freedom to be compress along the direction of propagation of the shock. It is therefore likely that flows along the directions parallel to the shock front can no longer be neglected and that the subsequent isochoric evolution predicted by the model (see Fig.~\ref{Fig-Trajectory-Main}) is somehow artificial. In this work, we thus assume that the model is reliable and can be integrated until the density reaches a fraction $\chi = 0.95$ of the final density $\densfin$. Such a criterion is found to prevent the integration of an unrealistic amount of postshock material and to provide reliable results on the column densities and the width of the line profiles which vary by less than a factor of four and less than a factor of two, respectively, for $\chi$ varying between 0.8 and 0.99.

Within this prescription, and for strong J-type shocks, the crossing time is equivalent to the isobaric cooling time. Indeed, for most of the models explored in this work, the crossing time is found to weakly depend on the strength of the magnetic field as long as $B_0 \leqslant 3$~$\mu$G and to scale (for $10 \leqslant V_S \leqslant 200$~\kms) as
\begin{equation} \label{Eq-cooltime1}
t_{\rm c} = 10^4\,\,{\rm yr}\,\, \left(\frac{\densini}{1\ \cc}\right)^{-1} \left(\frac{V_S}{100\ \kms}\right)^{-3} \,\, {\rm for}\,\, V_S \leqslant 100\,\,\kms
\end{equation}
and
\begin{equation} \label{Eq-cooltime2}
t_{\rm c} = 10^4\,\,{\rm yr}\,\, \left(\frac{\densini}{1\ \cc}\right)^{-1} \left(\frac{V_S}{100\ \kms}\right)^{2} \,\, {\rm for}\,\, V_S > 100\,\,\kms,
\end{equation}
as expected from the dependence of the cooling rate on the gas kinetic temperature. This crossing time corresponds to a shock size that ranges between $\sim 10^{-3}$~pc and $\sim 1$~pc over the entire grid of models.

\subsection{Chemistry and excitation of CI}

\begin{figure}[!h]
\begin{center}
\includegraphics[width=9.0cm,trim = 2.0cm 3.7cm 1.8cm 0.5cm, clip,angle=0]{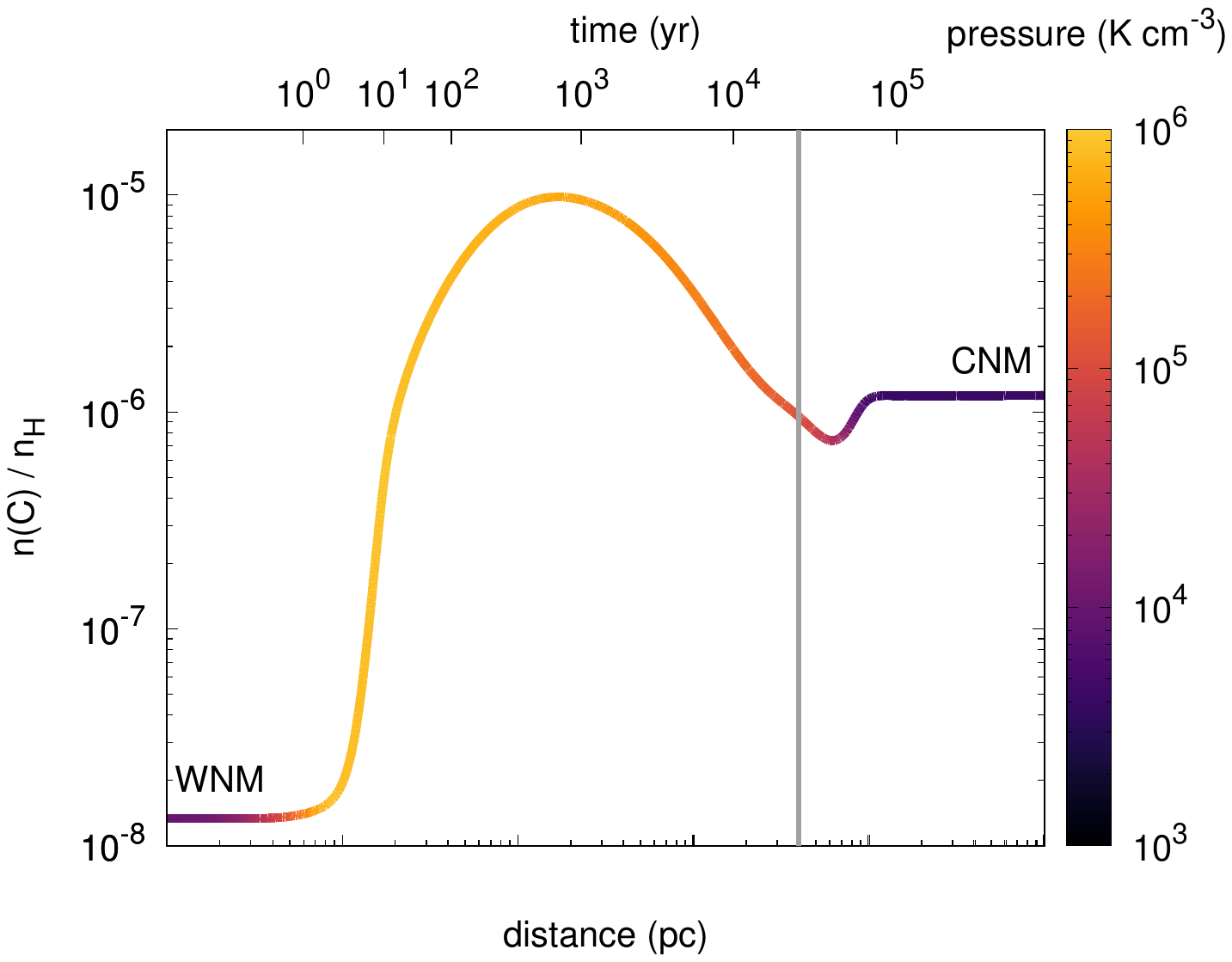}
\includegraphics[width=9.0cm,trim = 2.0cm 1.5cm 1.8cm 2.7cm, clip,angle=0]{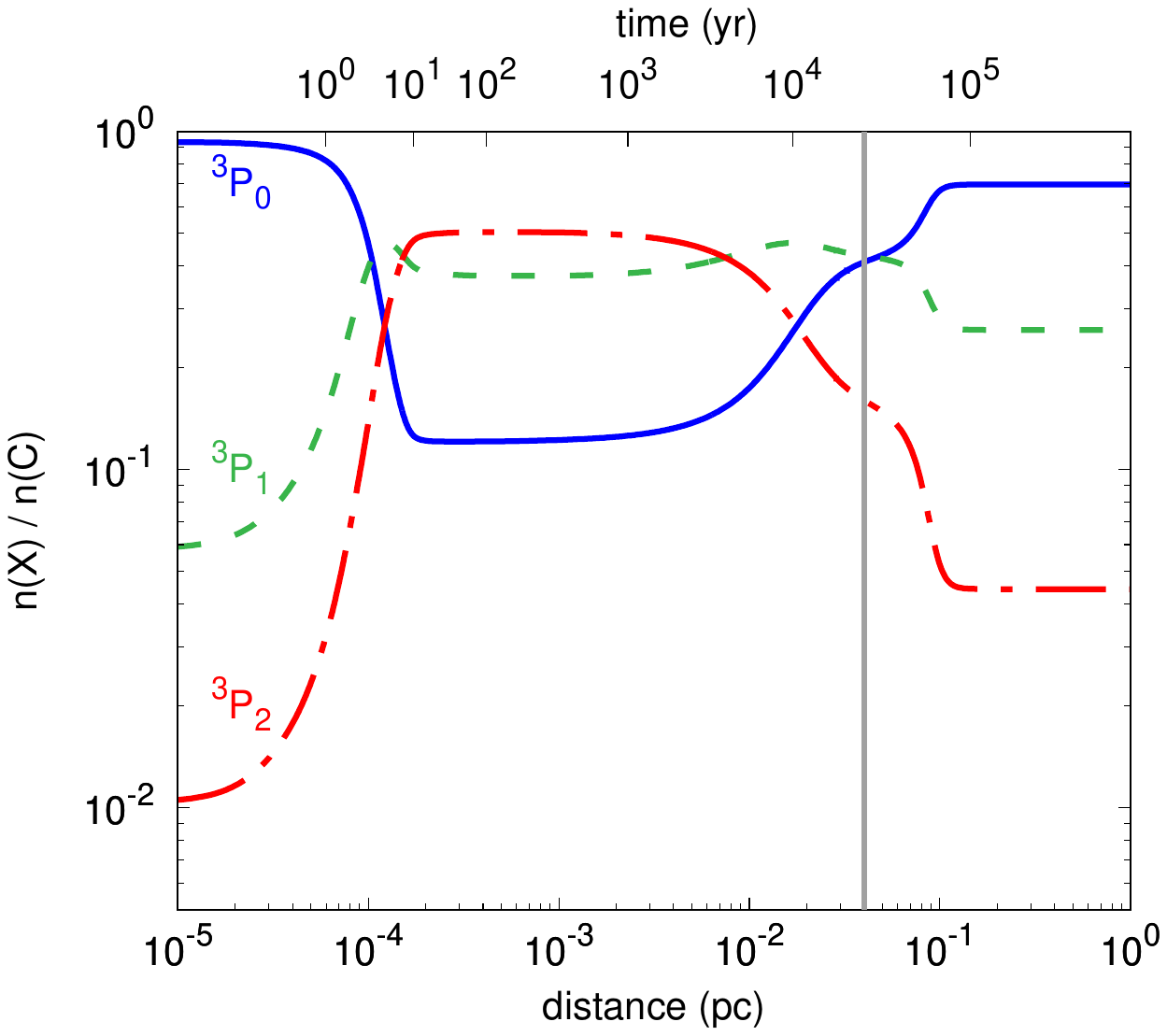}
\caption{Chemical and excitation profiles of neutral carbon across the standard model. {\it Top panel}: relative abundance of neutral carbon as a function of the distance (bottom axis) or time (top axis) in the shock. The curve is color-coded according to the value of the thermal pressure along the trajectory. {\it Bottom panel}: populations of the three fine structure levels of C as a function of the distance (bottom axis) or time (top axis) in the shock. The grey vertical line in both panel indicate the distance and time at which the gas reaches a fraction $\chi = 0.95$ of the final density $\densfin$.}
\label{Fig-Carbon-excit}
\end{center}
\end{figure}

The evolutions of the abundance of neutral carbon and of the populations of its three fine structure levels obtained in the standard model are shown in Fig.~\ref{Fig-Carbon-excit}. As the gas is converted from the WNM to the CNM, the compression induced by the shock wave  leads to an increase of the recombination of C$^+$, hence of the relative abundance of C. Unexpectedly, shocks induce an overproduction of neutral carbon, compared to the relative abundances found in the WNM and the CNM, that occurs before the thermal pressure decreases. This fundamental feature is explained by comparing the chemical timescales to the shock cooling time. Because interstellar shocks partly or even fully ionize neutral hydrogen, the recombination of C$^+$ is dominated by radiative and dielectronic recombinations\footnote{The fact that radiative and dielectronic recombinations dominate over the recombinations on grains and PAHs is reassuring. It implies that the destruction of large grains and PAHs, which starts to be important for shocks propagating at $V_S > 70$~\kms\ \citep{Jones1996,Micelotta2010a,Micelotta2010b} and is currently not taken into account in the model, has no impact on the results presented in this work.} with a characteristic timescale \citep{Badnell2003,Badnell2006}
\begin{equation} \label{Eq-recomb-time}
t_{\rm r} \sim 4 \times 10^2\,\,{\rm yr}\,\, \left(\frac{\densini}{1\ \cc}\right)^{3/2} \left(\frac{V_S}{80\ \kms}\right)\ \left(\frac{B_0}{1\ \mu{\rm G}}\right)^{-1}.
\end{equation}
The destruction of C, in the part of the trajectory which contributes the most to the column density of neutral carbon, is dominated by photoionization with a characteristic timescale \citep{Heays2017}
\begin{equation} \label{Eq-photon-time}
t_{\gamma} \sim 10^2\,\,{\rm yr}\,\, \left(\frac{G_0}{1}\right)^{-1}.
\end{equation}
These two scalings (Eqs.~\ref{Eq-recomb-time} and \ref{Eq-photon-time}) imply that the time required for neutral carbon to reach chemical equilibrium is always, at least, one hundred times smaller than the isobaric cooling time (see Eqs.~\ref{Eq-cooltime1} and \ref{Eq-cooltime2}). It follows that interstellar shock waves generate large column densities of neutral carbon with excitation conditions typical of a gas at high thermal pressure.

\begin{figure}[!h]
\begin{center}
\includegraphics[width=9.0cm,trim = 1.0cm 3.0cm 0.8cm 1.5cm, clip,angle=0]{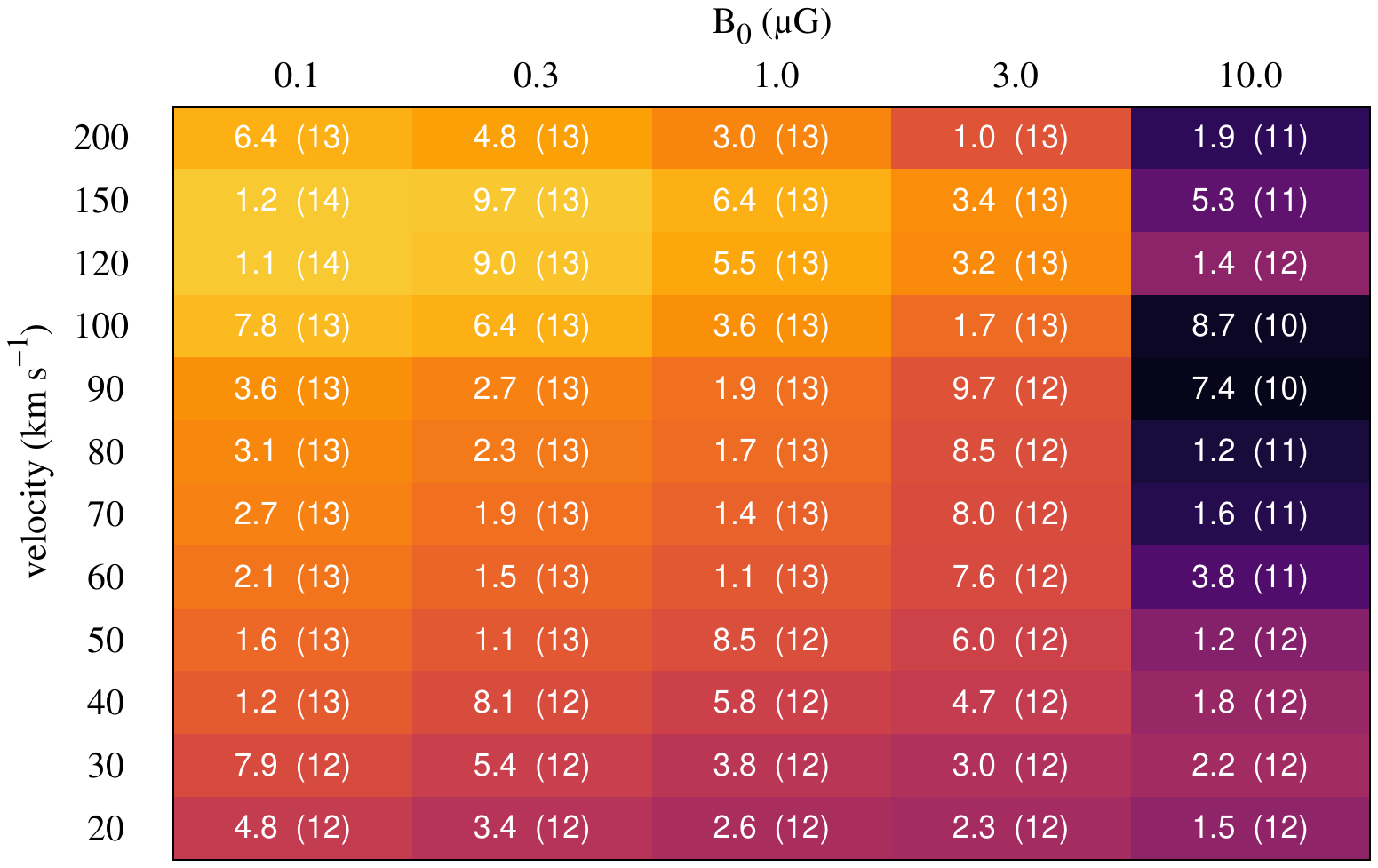}
\caption{Colored table of the column densities of neutral carbon (in cm$^{-2}$) computed across shocks for different velocities and different strength of the preshock magnetic field. The column densities are integrated in the direction perpendicular to the shock front over gas with a proton density $\dens \leqslant 0.95\,\,\densfin$. All other parameters are set to their standard values (see Table~\ref{Tab-main}). Numbers in parenthesis are powers of 10.}
\label{Fig-column-shocks-grid}
\end{center}
\end{figure}

\begin{figure}[!h]
\begin{center}
\includegraphics[width=9.0cm,trim = 3.5cm 2.7cm 2.0cm 2.0cm, clip,angle=0]{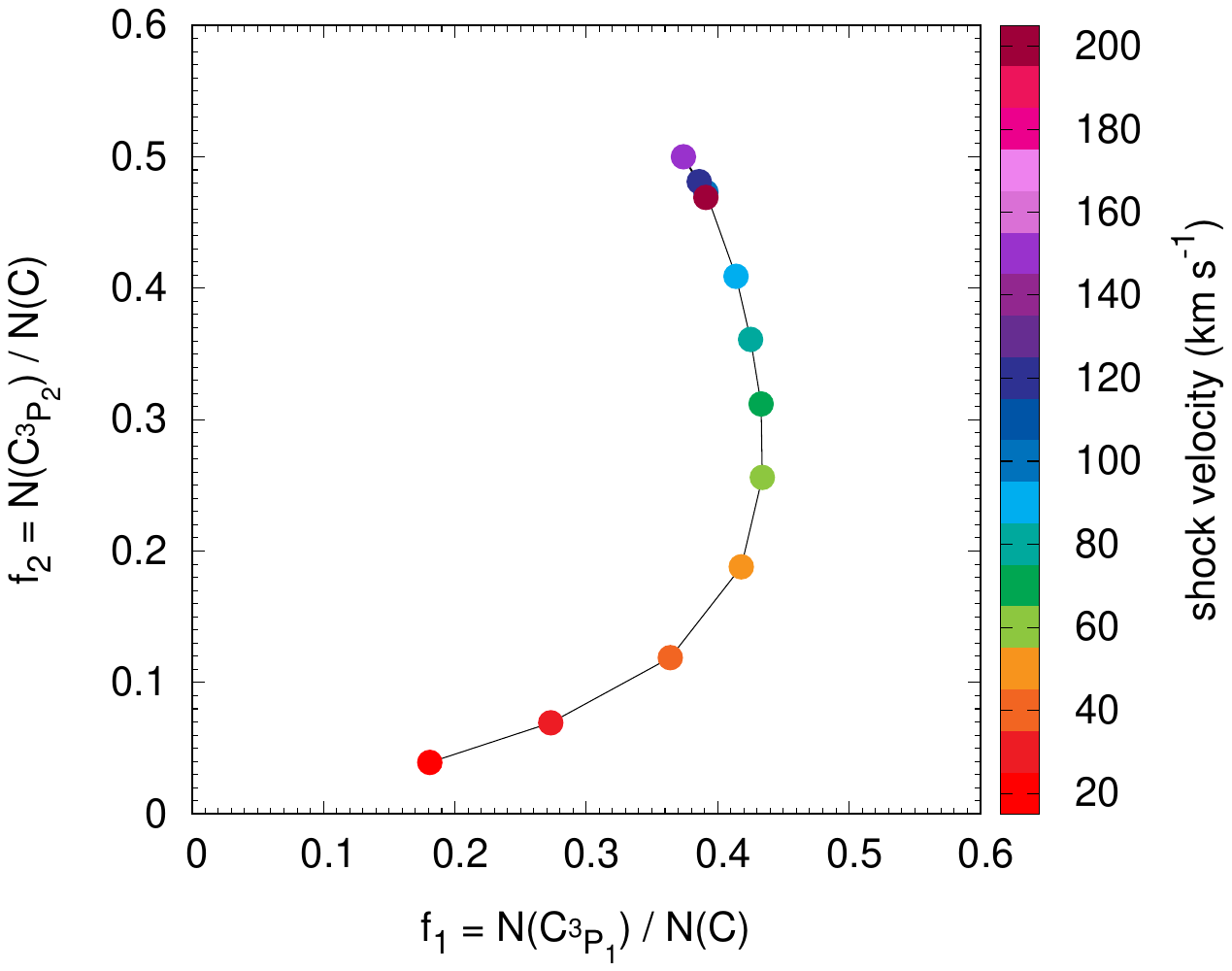}
\caption{$f_1$ and $f_2$ column density ratios computed across shocks at different velocities. The column densities are integrated in the direction perpendicular to the shock front over gas with a proton density $\dens \leqslant 0.95\,\,\densfin$. The color of each point indicates the velocity of the shock. All other parameters are set to their standard values (see Table~\ref{Tab-main}).}
\label{Fig-f1f2-shock}
\end{center}
\end{figure}

To illustrate this result, we display in Figs.~\ref{Fig-column-shocks-grid} and \ref{Fig-f1f2-shock} the total column densities of neutral carbon in shocks for different velocities and transverse magnetic field and the $f_1$ and $f_2$ column density ratios obtained in shocks at different velocities. In both figures, the column densities are calculated from the integration of the local densities, in the direction perpendicular to the shock front, up to the point where the proton density reaches a fraction $\chi = 0.95$ of its final value $\densfin$ (see, for instance, the grey vertical line in Fig.~\ref{Fig-Carbon-excit}). Fig.~\ref{Fig-column-shocks-grid} shows that interstellar shocks propagating in the WNM produce column densities of neutral carbon that range between a few $10^{12}$ and $10^{14}$~cm$^{-2}$ for $B_0 \leqslant 3$~$\mu$G. This range of values obtained in single shocks is in remarkable agreement with the column densities of neutral carbon derived along the 14 lines of sight where the gas at high thermal pressure was identified as a separated velocity component (see Table.~\ref{Tab-obs-prof}). 

As shown in Fig.~\ref{Fig-f1f2-shock}, the excitation conditions of C in interstellar shocks trace the postshock thermal pressure and are therefore highly dependent on the shock velocity (see Eq.~\ref{Eq-postpress}). For the models displayed here, the $f_1$ and $f_2$ column density ratios are found to cover a large range of values with ($f_1$, $f_2$) combinations of (0.18, 0.04) and (0.38, 0.5) for $V_S=20$ and 150~\kms\ respectively. When compared with Fig.~\ref{Fig-Decomp}, the combinations of $f_1$ and $f_2$ ratios shown in Fig.~\ref{Fig-f1f2-shock} are found to be either those required to explain individual lines of sight or those required to interpret the lines of sight as combinations of environments at low and high thermal pressure. All these results indicate that shocks propagating in the WNM could be a natural solution to the presence of neutral carbon at high pressure in the local diffuse ISM, provided that they produce kinematic signatures similar to the observations.

\subsection{Line profiles}

The kinematic signature of shocks are explored by calculating synthetic spectra of the column density of neutral carbon in its second excited level across shocks in the direction parallel to the direction of propagation. The velocity dependent column density spectrum writes
\begin{equation}
N({\rm C_{^3P_2}})(\upsilon) = \int_0^L n({\rm C_{^3P_2}})\,\phi(\upsilon)\, dz,
\end{equation}
where $n({\rm C_{^3P_2}})$ is the local density of neutral carbon in the second excited level, $\phi(\upsilon)$ is the local line profile, $z$ is the direction of propagation of the shock, and the integral is performed up to the distance $L$ where the local proton density $\dens=0.95\densfin$. The local line profile is assumed to be Gaussian and is computed as
\begin{equation}
\phi({\upsilon}) = \frac{1}{\sqrt{2\pi}\sigma_\upsilon} {\rm exp}\left(-\frac{1}{2} \left[\frac{\upsilon-\upsilon_0}{\sigma_\upsilon}\right]^2 \right),
\end{equation}
where $\upsilon_0$ is the local velocity of the gas along the direction of propagation of the shock and $\sigma_\upsilon$ is the local 1D velocity dispersion
\begin{equation}
\sigma_\upsilon = \left(\frac{kT}{m}+\sigma_{\rm tur}^2 \right)^{1/2},
\end{equation}
set by the local kinetic temperature, $T$, the mass of neutral carbon, $m$, and the micro-turbulent velocity dispersion, $\sigma_{\rm tur}$. For simplicity, and because there are no information regarding the amplitude of the micro-turbulence generated in the postshock gas, $\sigma_{\rm tur}$ is set to 0.

\begin{figure}[!h]
\begin{center}
\includegraphics[width=9.0cm,trim = 1.0cm 3.0cm 0.8cm 1.5cm, clip,angle=0]{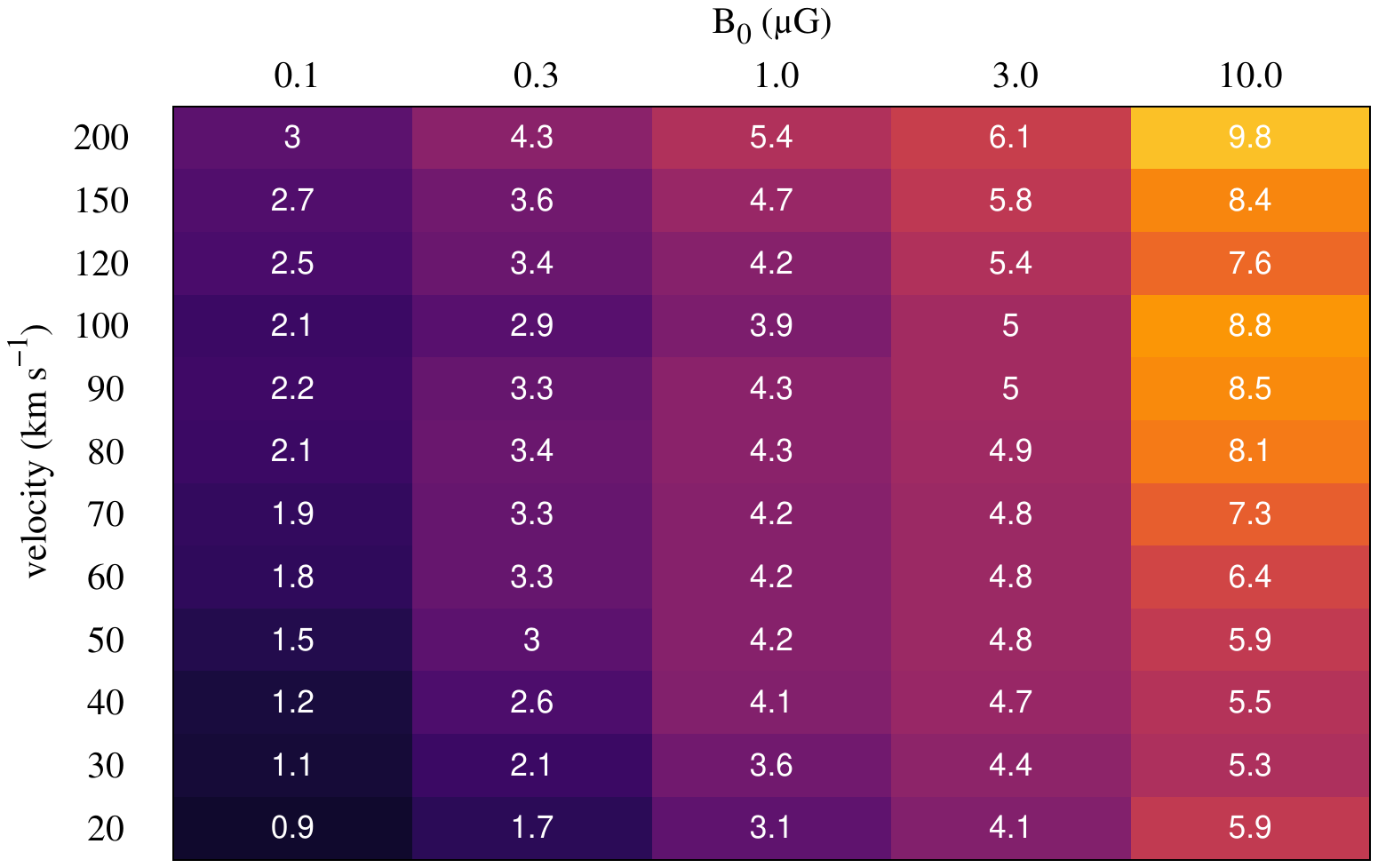}
\caption{Colored table of the FWHM (in \kms) of the absorption lines originating from the $^3P_2$ level of neutral carbon computed across shocks for different velocities and different strength of the preshock magnetic field. The FWHM are obtained by fitting Gaussian profiles on the velocity profiles predicted by the shock model in the direction perpendicular to the shock front taking only into account gas with a proton density $\dens \leqslant 0.95\,\,\densfin$. All other parameters are set to their standard values (see Table \ref{Tab-main}).}
\label{Fig-width-shocks-grid}
\end{center}
\end{figure}

The calculations applied to all models show that the synthetic spectra of interstellar shocks display Gaussian line profiles. This is due to the fact that the column density of neutral carbon is mostly produced in the postshock medium, where the velocity of the gas is roughly constant. The line profile integrated over the shock structure is therefore dominated by the thermal broadening set by the evolution of the kinetic temperature during the isobaric cooling of the postshock gas, with a weak, yet non negligible, contribution of the velocity gradient. To quantify further this feature, the synthetic spectra are fitted with a Gaussian component using the curve fit optimization algorithm provided by the SciPy python library. The result of this procedure is shown in Fig.~\ref{Fig-width-shocks-grid} which displays the FWHM of the column density spectra of shocks for different velocities and transverse magnetic field.

As the intuition dictates, the dependances of the predicted FWHM on the shock velocity, the transverse magnetic field, and the preshock density, are found to roughly follow the dependancies expected from the square root of the ratio of the postshock thermal pressure (Eq.~\ref{Eq-postpress}) and the final density (Eq.~\ref{Eq-densfin}), except for the dependance on $B_0$ which is somehow shallower. Keeping in mind that the predictions could be affected by the turbulent motions generated in the wake of interstellar shocks, the FWHM are found to range from 1 to 6~\kms, for a transverse magnetic field $B_0 \leqslant 3$~$\mu$G, in remarkable agreement with the range of FWHM derived from the observations in velocity components solely associated to the high pressure gas (see Table~\ref{Tab-obs-prof}). This result, combined with those presented in the previous section, implies that shocks propagating in the WNM meet all the criteria required to explain the presence of neutral carbon at high pressure. It also clearly rule out shocks propagating in highly magnetized WNM ($B_0>3$~$\mu$G). The question now is whether this possible scenario is plausible or not.

\section{Dissipation rate} \label{Sect-grid}

\subsection{Identification of individual shocks}

The fact that the line profiles predicted by the model are Gaussian with linewidths similar to those observed for individual velocity components at high pressure suggests that each of these velocity component is associated to a single shock. The number of velocity components clearly identified as gas at high pressure also suggest that each line of sight observed by \citetalias{Jenkins2011} intercept only one or a few shocks propagating in the WNM and not a collection of tens or hundreds of shocks. It would therefore be tempting to use the predictions given in the previous section to analyse the lines of sight observed by \citetalias{Jenkins2011} and identify the properties of individual shocks.

For instance, the column density spectrum of HD~210839 (see Fig.~\ref{Fig-HD210839}) displays two velocity components at high pressure, centered at $V=-34$ and $-31$~\kms, with FWHM of 2.6 and 2.9~\kms, and total column densities of neutral carbon of $7.1 \times 10^{13}$ and $5 \times 10^{13}$~cm$^{-2}$ (see Table~\ref{Tab-obs-prof}). The results given in Figs.~\ref{Fig-column-shocks-grid} and \ref{Fig-width-shocks-grid} indicate that the observations could be explained by two shocks propagating along the line of sight at velocities between 80 and 100~\kms\ in a medium with a preshock density of 1~\cc\ and a transverse magnetic field between 0.1 and 0.3~$\mu$G. There are, however, two problems with this approach. The first problem is that such a solution is not unique. Although not shown in Figs.~\ref{Fig-column-shocks-grid} and \ref{Fig-width-shocks-grid}, equally acceptable solutions could be found with different sets of shock parameters ($V_S$, $\densini$, and $B_0$). Moreover, this method depends on the angle of inclination of the shock along the line of sight, which affects the predicted column density, and on the amplitude of turbulent motions in the post-shock flow, which may affect the width of the line profiles, two unknown quantities. The second problem is that this method can only be applied on the 14 lines of sight where velocity components at high thermal pressure are identified and not on the entire sample of \citetalias{Jenkins2011}.

The various degeneracies cited above implies that the sole observations of the excited levels of neutral carbon are insufficient to identify the exact properties of the shock and of the medium in which they propagate. The search for additional observational tracers  will be carried in forthcoming papers. Short of these additional tracers, we devise, in the following, a method to analyze every lines of sight and derive a more meaningful and robust quantity: the distribution of the dissipation rate of mechanical energy required to explain the observations.

\subsection{Decomposition method}

\begin{figure}[!h]
\begin{center}
\includegraphics[width=9.0cm,trim = 1.5cm 1cm 0.5cm 2.0cm, clip,angle=0]{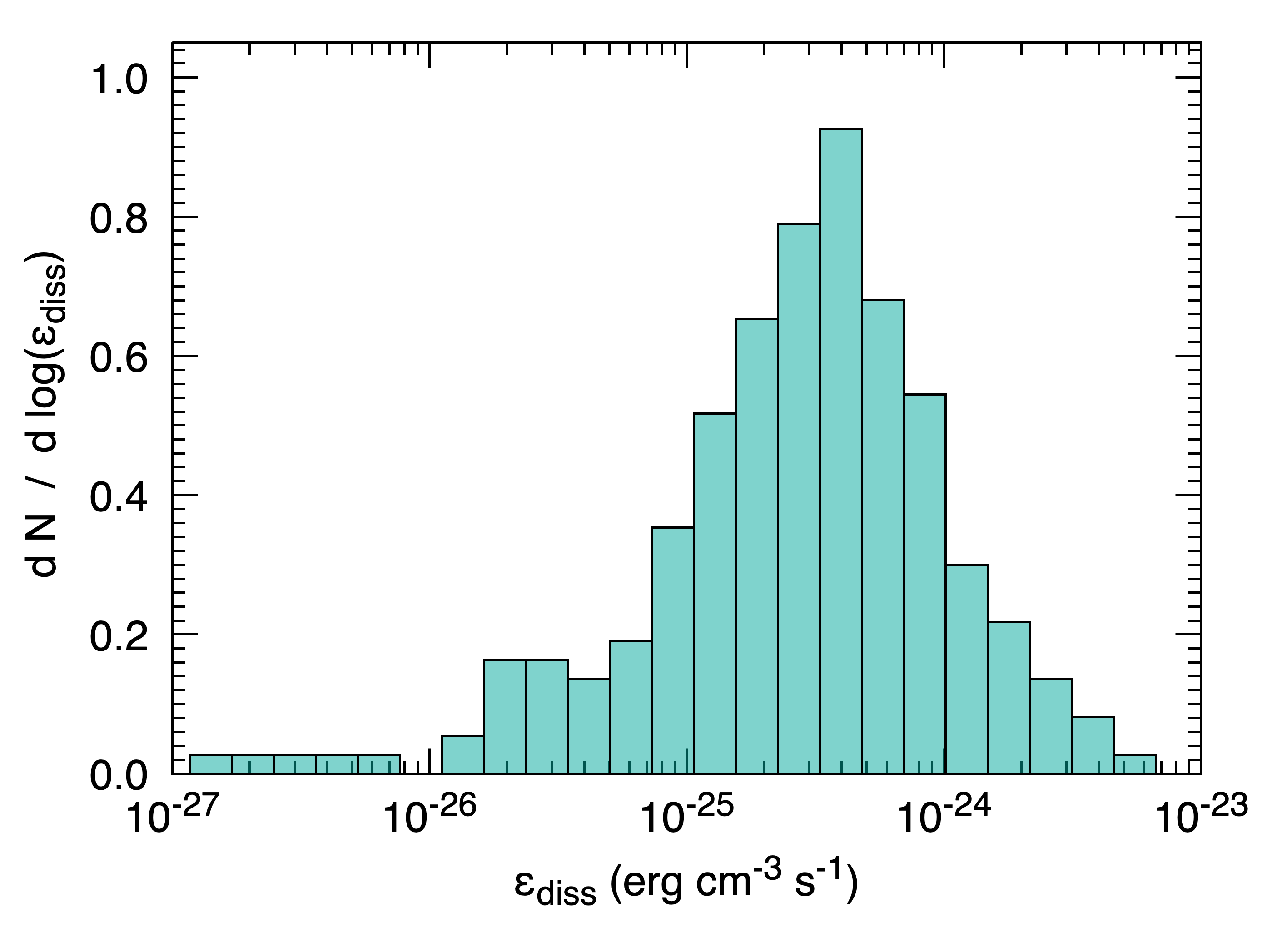}
\caption{Probability distribution function of the dissipation rate required to reproduce the column density of carbon at high pressure along the $N$ lines of sight of \citetalias{Jenkins2011} assuming that all the shocks intercepted by these lines of sight are identical and defined by the parameters of the standard model: $\densini = 1$~\cc, $B_0 = 1$~$\mu$G, and $V_S = 80$~\kms. The probability distribution function is normalized so that its integral is equal to one.}
\label{Fig-eps}
\end{center}
\end{figure}

To illustrate the method, let's consider first a simple case and assume that all the shocks propagating in the WNM have the same physical properties: the same preshock density, $\densini$, shock velocity, $V_S$, and transverse magnetic field, $B_0$. If those parameters are fixed, the $f_1$ and $f_2$ column density ratios integrated across this shock are known and given by the model (see Fig~\ref{Fig-f1f2-shock}). If the gas at high thermal pressure corresponds to shocked material, the decomposition procedure of \citetalias{Jenkins2011} (see Sect.~\ref{Sect-decomp-method} and Fig.~\ref{Fig-Decomp}) can be applied adopting the ($f_1$, $f_2$) combination predicted by the model for the gas at high thermal pressure instead of an arbitrary combination (see Sect.~\ref{Sect-decomp-method}). This decomposition procedure provides the fraction of gas at high thermal pressure and the associated column density of neutral carbon, $N^{\rm o}_{\rm high}({\rm C})$, along any given line of sight. Therefore, the number of shocks required to explain each observation simply writes
\begin{equation} \label{Eq-nbshock}
\mathcal{N}_{\rm shk} = \frac{N^{\rm o}_{\rm high}({\rm C})}{N^{\rm m}({\rm C})},
\end{equation}
where $N^{\rm m}({\rm C})$ is the column density of carbon predicted by the model (see Fig.~\ref{Fig-column-shocks-grid}). It follows that the rate of dissipation of mechanical energy (in~\eccs), averaged over the length of the neutral gas, writes
\begin{equation}
\varepsilon_{\rm diss} = \frac{1}{2}\,\mu\,\densini V_S^3\,\frac{\mathcal{N_{\rm shk}}}{l_{\rm los}\varphi},
\end{equation}
where $\mu$ is the mean particle mass of the atomic gas (in~g) and $\varphi$ is the fraction of volume occupied by the ionized gas along the line of sight set to the conservative value of 0.5 (see Sect.~5.3 of \citealt{Bellomi2020}). $l_{\rm los}$ is the length of diffuse gas (ionized or not) intercepted by the line of sight and estimated with Eq.~\ref{Eq-length}.

The result of this methodology is shown in Fig.~\ref{Fig-eps} which displays the probability distribution function of the logarithm of the dissipation rate obtained after applying the decomposition procedure to the entire observational sample, assuming that all the shocks are identical and correspond to the standard model ($\densini = 1$~\cc, $B_0 = 1$~$\mu$G, and $V_S = 80$~\kms). Under this strong assumption, the dissipation rate is found to cover more than three orders of magnitude, with a median of $3 \times 10^{-25}$~\eccs, and a dispersion of about 0.4 dex (corresponding to a factor of 2.5). This distribution weakly depends on the prescription adopted in Sect.~\ref{Sect-obs-pos} to compute the length of the intercepted diffuse material (see Eq.~\ref{Eq-length}). This is because most of the lines of sight are located at low Galactic latitude where $l_{\rm los}$ is close or equal to the distance of the background source (see Fig.~\ref{Fig-Obs-pos}).

\subsection{Impact of the parameters}

\begin{figure*}[!h]
\begin{center}
\includegraphics[width=19.0cm,trim = 1.5cm 2.0cm 0.5cm 0.0cm, clip,angle=0]{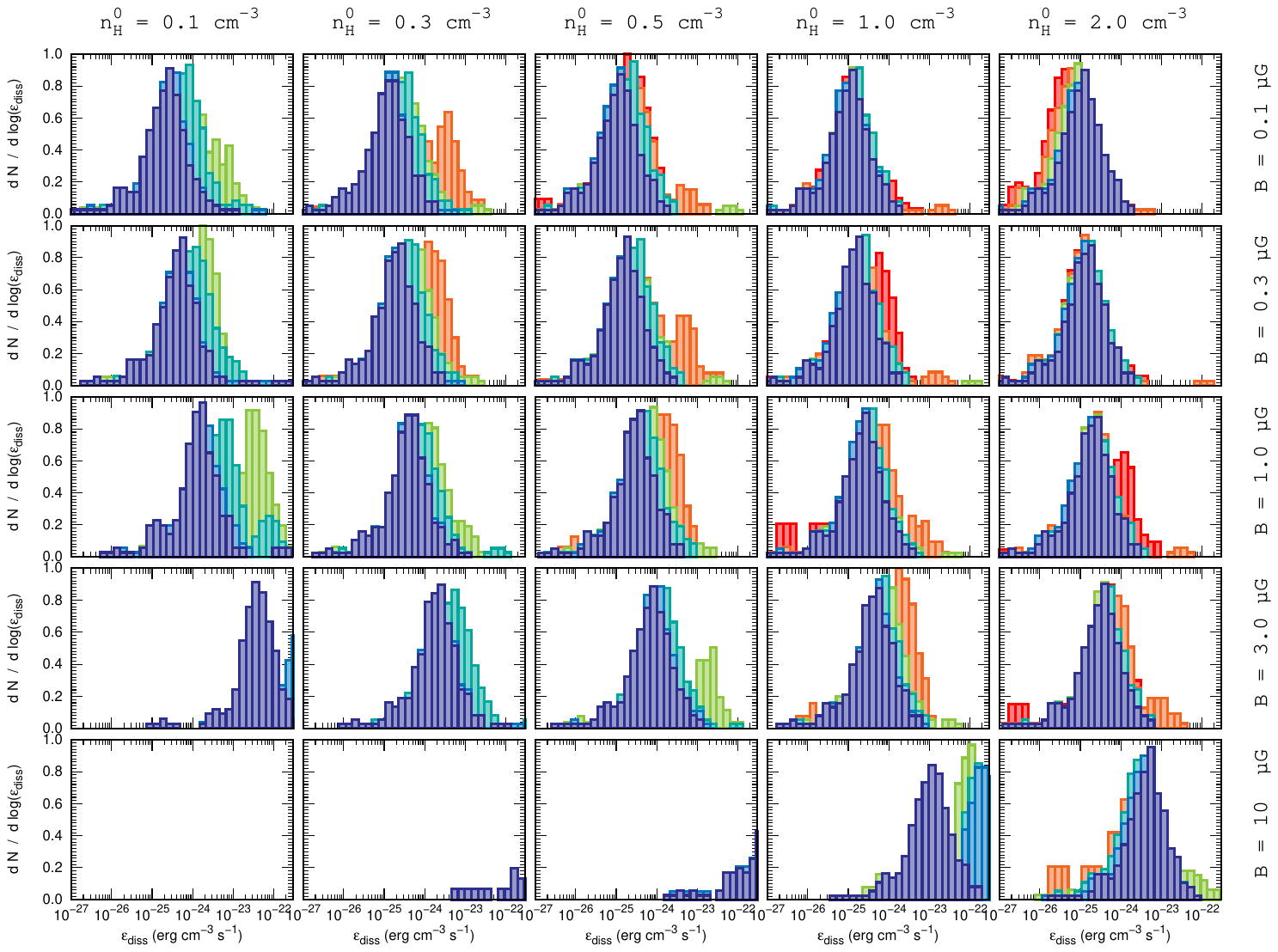}
\caption{Same as Fig. \ref{Fig-eps} for a preshock density varying between 0.1 and 2~\cc\ (from left to right), a strength of the preshock magnetic field varying between 0.1 and 10 $\mu$G (from top to bottom) and six values of the shock velocity: 20 (red), 40 (orange), 60 (green), 80 (turquoise), 100 (light blue), and 120 (dark blue) \kms. For coherence, the color code is the same than that used in Fig. \ref{Fig-f1f2-shock}. In many cases, only the last distribution (corresponding to $V_S = 120$~\kms) is visible, because those obtained at lower velocities are identical and hidden behind it. The bottom left panels are empty because the distributions fall outside the range of dissipation rates displayed.}
\label{Fig-eps-grd}
\end{center}
\end{figure*}

\begin{figure}[!h]
\begin{center}
\includegraphics[width=9.0cm,trim = 1.0cm 3.0cm 1.0cm 1.5cm, clip,angle=0]{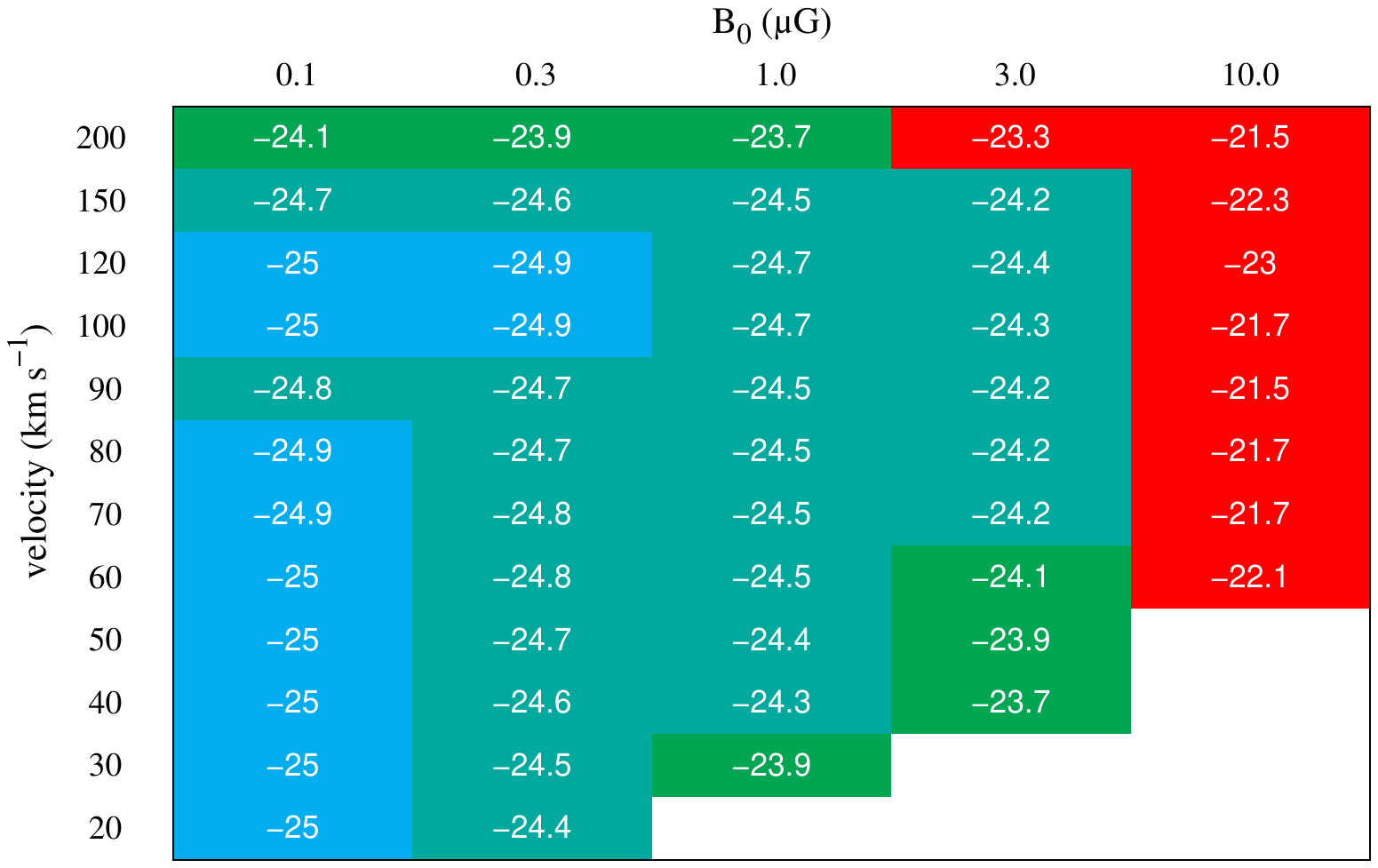}
\caption{Colored table of the median value of the dissipation rate (see Fig. \ref{Fig-eps-grd}) as a function of the shock velocity and the strength of the magnetic field for $\densini=1$~\cc. All the other parameters are set to their standard values (see Table~\ref{Tab-main}). Models with a median dissipation rate within a factor of two of that obtained with the standard model are shown in turquoise. Models with a median dissipation rate between two and four times larger (resp. lower) than that obtained with the standard model are shown in green (resp. blue). All other models are shown in red.}
\label{Fig-eps-grd-med}
\end{center}
\end{figure}

Although interesting, the distribution of dissipation rates obtained in Fig.~\ref{Fig-eps} has a limited value because it is derived assuming that all the shocks propagating in the WNM are identical. Such an assumption is obviously not realistic. Even if the density of the WNM in the local ISM is known to follow a log-normal distribution around a mean value of about 0.7~\cc\ \citep{Marchal2021a}, shocks are expected to occur in a variety of warm neutral environments with different densities and magnetization. Moreover, the probable driving sources of interstellar shocks (see Sect.~\ref{Sect-disc}) imply that the observed lines of sight necessarily intercept shocks with different velocities. These various distributions of physical conditions need to be taken into account to derive a meaningful distribution of dissipation rates.

The impacts of the model parameters on the decomposition procedure described in the previous section are shown in Fig.~\ref{Fig-eps-grd} which displays the distributions of the logarithm of the dissipation rate obtained for various sets of preshock density, transverse magnetic field, and shock velocity. As shown in Fig.~\ref{Fig-f1f2-shock}, the ($f_1$, $f_2$) combination highly depends on the shock velocity. In particular, shocks at low velocity lead to $f_1$ and $f_2$ ratios which are not sufficient to perform the decomposition of all the lines of sight. For $\densini=1$~\cc\ and $B_0 = 1$~$\mu$G, we find that shocks with velocities larger than 30-40~\kms\ are required to allow the decomposition of most of the observational sample (see Figs.~\ref{Fig-Jenkins} and \ref{Fig-f1f2-shock}). Each distribution displayed in Fig.~\ref{Fig-eps-grd} is thus computed and normalized taking only into account the lines of sight which can be decomposed for the associated set of shock parameters ($V_S$, $\densini$, and $B_0$).

Fig.~\ref{Fig-eps-grd} reveals two fundamental and surprising results. Firstly, the logarithm of the dissipation rate almost systematically follows a Gaussian distribution (plus a tail at low energy) with a dispersion of $\sim 0.4$, regardless of the model parameter. Secondly, most of the distributions shown in Fig.~\ref{Fig-eps-grd} appear to be superimposed. This last result is quantified further in Fig.~\ref{Fig-eps-grd-med}, which displays the median value of these distributions for $\densini=1$~\cc. As long as $B_0 \leqslant 3$~$\mu$G, the median dissipation rate is found to weakly depend on the model parameters and to vary by only a factor of $\sim 10$ for input fluxes of kinetic and magnetic energies varying by a factor of $\sim 10\,000$. For shock velocities between $\sim 40$ and $\sim 150$~\kms\ (turquoise and blue regions in Fig.~\ref{Fig-eps-grd-med}), the median dissipation rate obtained for $\densini=1$~\cc\ varies by a factor of $\sim 6$, which corresponds to only twice the dispersion of the distribution.

This weak dependence of the dissipation rate on the model parameters has no simple analytical form and basically results from two antagonistic effects. On the one side, larger velocities and preshock densities lead to higher column densities of neutral carbon, $N^{\rm m}({\rm C})$, with higher $f_2$ ratios, hence lower $N_{\rm high}^{\rm o}({\rm C})$. The number of shocks required to explain a given line of sight therefore drastically decreases when the velocity or the density increases 
(see Eq.~\ref{Eq-nbshock}). On the other side, shocks with larger velocities or preshock densities have larger individual dissipation rates. These antagonistic effects compensate each other almost perfectly, leading to a stable distribution of dissipation rates over a wide range of parameters.

The result displayed in Figs.~\ref{Fig-eps-grd} and \ref{Fig-eps-grd-med} is one of the main outcome of this work. It implies that the dissipation rate required to explain the observations of \citetalias{Jenkins2011} and the presence of carbon at high pressure is roughly independent of the exact nature of the shocks along the lines of sight and the medium in which they propagate. It also implies that the distribution displayed in Fig.~\ref{Fig-eps} is more meaningful than originally thought because a similar distribution would be obtained assuming a convoluted distribution of shock velocities, preshock densities, and transverse magnetic field. It finally means that the validity of this scenario can be estimated by considering the possible driving sources of interstellar shocks in the WNM and compare the associated distributions of dissipation rates with Fig.~\ref{Fig-eps}.

\section{A scenario of shocks driven by supernovae} \label{Sect-disc}

\subsection{Origin of high velocity shocks}

As shown in the previous sections, the presence of neutral carbon at high pressure could be explained by shocks propagating at velocities $V_S \gtrsim 30$~\kms\ in a WNM environment with a typical density, $\densini \geqslant 0.3$~\cc, and a magnetic field strength, $B_0 \leqslant 3$~$\mu$G. In theory, shocks with such velocities could be driven by a variety of phenomena in the local ISM including stellar winds, bipolar outflows, supernovae explosions, or the infall of matter onto the Galactic disk. The relative contributions of these events depend, however, on their individual rates, the deceleration time of the shock they produce, and the volume of the ISM they affect. Both analytical studies and numerical simulations show that supernovae explosions should dominate the injection rate of mechanical energy in the Milky Way (e.g. \citealt{Norman1996, Brucy2020a}). We therefore discuss here the scenario where shocks propagating in the WNM are driven by the expansion of SNRs in the diffuse phase of the local ISM.

\subsection{Expansion of supernova remnants}

To estimate the physical properties of shocks driven by supernova remnants, we consider the classical picture of a supernovae explosion with an ejecta mass, $M_{\rm ej}$, and an initial kinetic energy, $E$, expanding in an homogeneous medium with a proton density $\densini$. In an idealized spherical expansion, the resulting supernova remnant is known to evolve through four different stages, designated as the free expansion stage (FE), the Sedov-Taylor stage (ST), the pressure-driven stage (PD), and the momentum conserving stage (MC) (e.g. \citealt{Cioffi1988a, Truelove1999a, Kim2015a}). During each of these stages, the radius, $R_B$, and the velocity, $V_B$, of the terminal blast wave can be approximated by the following expressions (e.g. \citealt{Draine2011, Vink2020a})
\begingroup\makeatletter\def\f@size{9}\check@mathfonts
\begin{equation}
R_B= 
\begin{dcases}
\phantom{1}3.1\,\,{\rm pc}\,\,\left(\frac{M_{\rm ej}}{M_\odot}\right)^{1/3} \,\,  \left(\frac{\densini}{n_0}\right)^{-1/3} \left(\frac{t}{t_{\rm ST}}\right) & \text{if } 0 < t \leqslant t_{\rm ST} \\
\phantom{1}3.1\,\,{\rm pc}\,\,\left(\frac{M_{\rm ej}}{M_\odot}\right)^{1/3} \,\,  \left(\frac{\densini}{n_0}\right)^{-1/3} \left(\frac{t}{t_{\rm ST}}\right)^{2/5} & \text{if } t_{\rm ST} < t \leqslant t_{\rm PD}\\
23.8\,\,{\rm pc}\,\, \left(\frac{E}{E_{51}}\right)^{0.29} \,\,  \left(\frac{\densini}{n_0}\right)^{-0.42} \left(\frac{t}{t_{\rm PD}}\right)^{2/7} & \text{if } t_{\rm PD} < t \leqslant t_{\rm MC}\\
46.0\,\,{\rm pc}\,\, \left(\frac{E}{E_{51}}\right)^{0.29} \,\,  \left(\frac{\densini}{n_0}\right)^{-0.42} \left(\frac{t}{t_{\rm MC}}\right)^{1/4} & \text{if } t_{\rm MC} < t \leqslant t_{\rm fade}
\end{dcases}
\end{equation}
\endgroup
and
\begingroup\makeatletter\def\f@size{9}\check@mathfonts
\begin{equation}
V_B= 
\begin{dcases}
10\,000\,\,{\rm \kms}\,\, \left(\frac{E}{E_{51}}\right)^{1/2} \,\,  \left(\frac{M_{\rm ej}}{M_\odot}\right)^{-1/2} & \text{if } 0 < t \leqslant t_{\rm ST} \\
\phantom{1}3\,988\,\,{\rm \kms}\,\, \left(\frac{E}{E_{51}}\right)^{1/2} \,\,  \left(\frac{M_{\rm ej}}{M_\odot}\right)^{-1/2}  \left(\frac{t}{t_{\rm ST}}\right)^{-3/5} & \text{if } t_{\rm ST} < t \leqslant t_{\rm PD}\\
\phantom{11\,}135\,\,{\rm \kms}\,\, \left(\frac{E}{E_{51}}\right)^{0.07} \,\,  \left(\frac{\densini}{n_0}\right)^{0.13}  \left(\frac{t}{t_{\rm PD}}\right)^{-5/7} & \text{if } t_{\rm PD} < t \leqslant t_{\rm MC}\\
\phantom{11\,1}26\,\,{\rm \kms}\,\, \left(\frac{E}{E_{51}}\right)^{0.07} \,\,  \left(\frac{\densini}{n_0}\right)^{0.13}  \left(\frac{t}{t_{\rm MC}}\right)^{-3/4} & \text{if } t_{\rm MC} < t \leqslant t_{\rm fade},
\end{dcases}
\end{equation}
\endgroup
where $E_{51} = 10^{51}$~erg and $n_0 = 1$~\cc\ are normalization factors, $t_{\rm ST}$, $t_{\rm PD}$, and $t_{\rm MC}$ are the onset times of the ST, PD, and MC stages, respectively, and $t_{\rm fade}$ is the fade away time of the supernova. The onset time of the Sedov-Taylor stage is set to ensure the continuity of the radius between the FE and the ST stages\footnote{This definition leads to an onset time slightly larger (by a factor 1.6) than that obtained when the swept up mass of interstellar matter is comparable to the ejecta mass.},
\begin{equation}
t_{\rm ST} = 0.303\,\,{\rm kyr}\,\, \left(\frac{E}{E_{51}}\right)^{-1/2} \,\, \left(\frac{M_{\rm ej}}{M_\odot}\right)^{5/6} \,\, \left(\frac{\densini}{n_0}\right)^{-1/3}.
\end{equation}
Following \citet{Draine2011}, the onset time of the pressure-driven stage is set to the moment when the integrated radiative losses of the hot gas become comparable to its thermal energy ($\Delta E_{\rm th} / E_{\rm th} = -1/3$),
\begin{equation}
t_{\rm PD} = 49.3\,\, {\rm kyr}\,\, \left(\frac{E}{E_{51}}\right)^{0.22} \,\, \left(\frac{\densini}{n_0}\right)^{-0.55}.
\end{equation}
The onset of the momentum conserving stage is hard to estimate \citep{Kim2015a}. To simplify we set it here to 10 times the onset time of the PD stage. The fade away time is finally taken as the moment where the terminal velocity falls bellow the 1D velocity dispersion of the WNM, set to 10~\kms,
\begin{equation} \label{Eq-fade}
t_{\rm fade} = 1.76\,\, {\rm Myr}\,\, \left(\frac{E}{E_{51}}\right)^{0.31} \,\, \left(\frac{\densini}{n_0}\right)^{-0.38}.
\end{equation}

\begin{figure}[!h]
\begin{center}
\includegraphics[width=9.0cm,trim = 13.0cm 3.5cm 1.0cm 4.5cm, clip,angle=0]{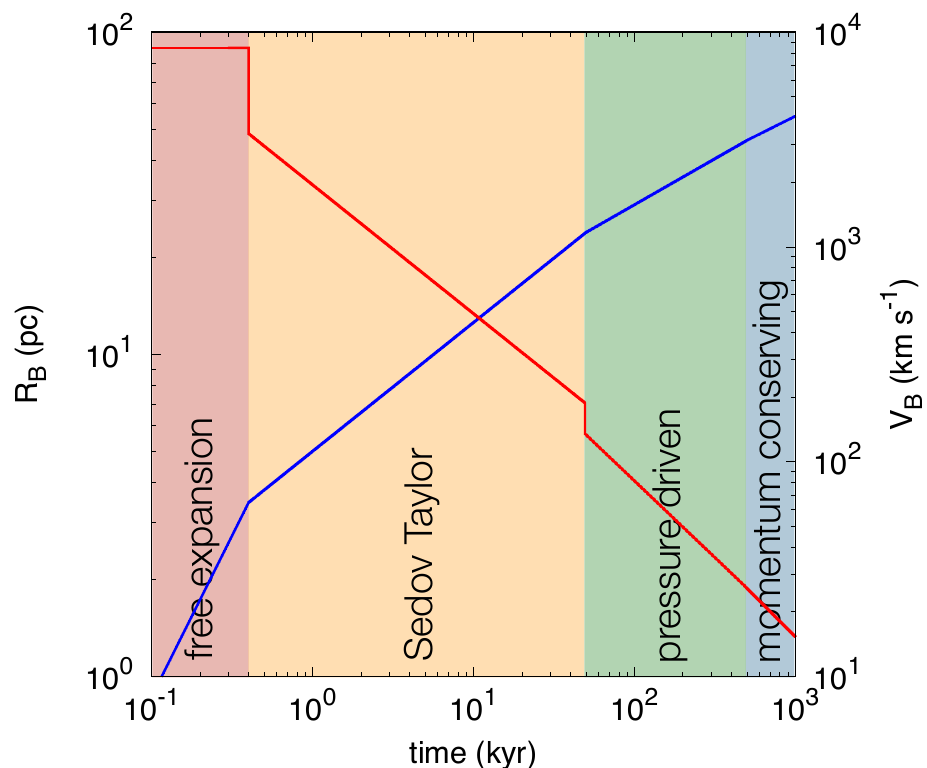}
\caption{Basic analytical descriptions of the evolutions of the radius (blue curve) and the terminal velocity (red curve) of a one-dimensional supernova remnant (e.g. \citealt{Draine2011,Vink2020a}), characterized by an ejecta mass $M_{\rm ej} = 1.4$~$M_\odot$ and an initial kinetic energy energy $E = 10^{51}$~erg, and expanding in an homogeneous medium with a density of 1~\cc. The idealized successive phases known as the free expansion stage, the Sedov-Taylor stage, the pressure-driven stage, and the momentum conserving (or snowplow) stage are highlighted in red, orange, green, and blue, respectively.}
\label{Fig-SNR-analytics}
\end{center}
\end{figure}

Fig.~\ref{Fig-SNR-analytics} displays the evolutions of the blast radius and velocity considering the prototypical case of a supernova explosion with $M_{\rm ej} = 1.4$~$M_\odot$, $E = 10^{51}$~erg, and $\densini = 1$~\cc\ \citep{Draine2011}. The discontinuities of the terminal velocity come from the idealized descriptions adopted above which neglect the deceleration of the blast wave by the surrounding environment at the end of the FE stage and the cooling of the hot shell at the end of the ST stage. Fig.~\ref{Fig-SNR-analytics} shows that a typical SNR produces a terminal shock with a velocity above $\sim 30$~\kms\ over long periods that last from a few tens to a few hundreds kyr and is therefore a likely source of intermediate and high velocity shocks in the WNM. Interestingly, the terminal velocity at the end of the Sedov-Taylor stage is $\sim 188$~\kms\ with small dependences on the model parameters. Since the blast deceleration timescale is smaller than the cooling timescale of the hot gas in the ST stage, this result implies that shocks driven by SNRs at velocities larger than $\sim 200$~\kms\ are not radiative \citep{Vink2012a} and cannot be treated at steady-state.

\subsection{Dissipation rate of a simplistic distribution of SNR}

\begin{figure}[!h]
\begin{center}
\includegraphics[width=8.0cm,trim = 1.0cm 1.5cm 19.5cm 0.3cm, clip,angle=0]{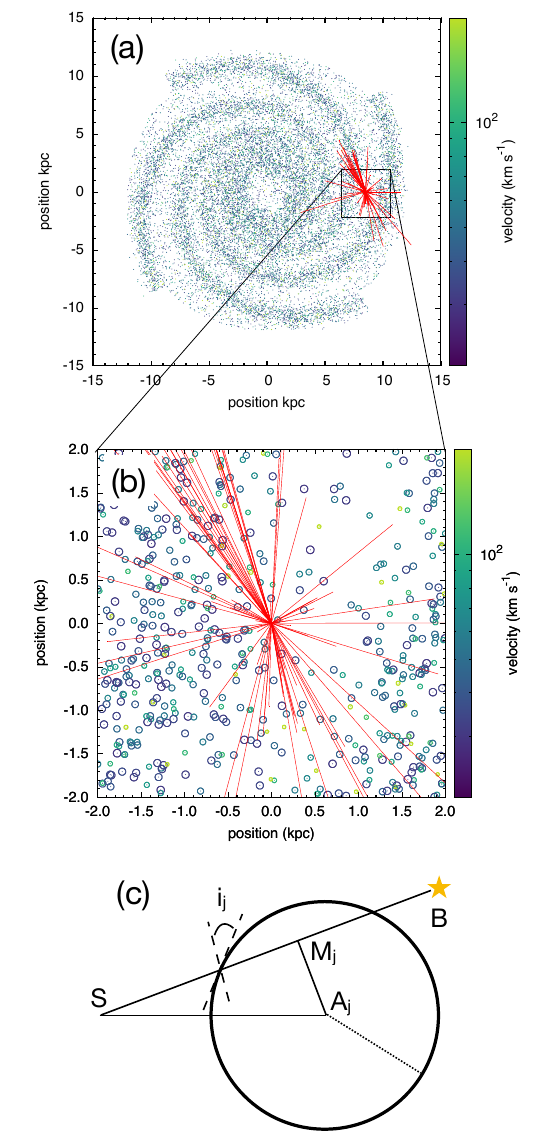}
\caption{Schematic view of the calculation of the dissipation rates expected from SNRs along the lines of sight observed by \citetalias{Jenkins2011}. A distribution of SNRs with terminal velocities between 30 and 200~\kms\ is drawn out of a simple description of supernovae explosions within the disk and the spiral arms of the Galaxy (panel a). The SNRs are color-coded according to their terminal velocities and their sizes correspond to the radius of the associated blast waves. The lines of sight observed by \citetalias{Jenkins2011} (red lines) are analysed to count the number of SNR surfaces they intercept and compute the dissipation rates induced by the associated shocks (panels b and c). This calculation is done taking into account the 3D spatial distribution of SNRs which is shown in projection in panels (a) and (b). Panel (c) displays a schematic view of a single SNR located at a position A$_j$. An observed line of sight connects the position of the sun, S, to the position of the background star, B. The number of surfaces crossed by the line of sight for this single SNR (0, 1, or 2) and their inclination angle i$_j$ depend on the position of the point M$_j$ along the SB line and its distance to A$_j$ compared to the SNR radius.}
\label{Fig-SNR-dist}
\end{center}
\end{figure}

Estimating the dissipation rate of mechanical energy induced by SNRs along the lines of sight observed by \citetalias{Jenkins2011} requires a description of the spatial distribution of supernovae in the Galaxy. To perform a back of the enveloppe calculation of the distribution of the dissipation rate, we adopt the following simplified picture. All supernovae are assumed to have the same ejecta mass, $M_{\rm ej} = 1.4$~$M_\odot$, and initial kinetic energy, $E = 10^{51}$~erg, and to expand in a medium defined by a single proton density, $\densini = 1$~\cc\ (see Fig.~\ref{Fig-SNR-analytics}). The rate of supernova explosions in the Galaxy, $k_{\rm SN}$, is set to 46~kyr$^{-1}$ \citep{Adams2013a}. In a permanent regime, the number of supernova remnants in the Galaxy therefore writes
\begin{equation}
\mathcal{N}_{\rm SNR} = k_{\rm SN}\,\,t_{\rm fade},
\end{equation}
where $t_{\rm fade}$ is the SNR fade away time (see Eq.~\ref{Eq-fade}). Based on the model of \citet{Adams2013a}, we assume that 70\% of the supernovae correspond to core collapses (SNII, SNIb, and SNIc) and are distributed along the spiral arms, and that 30\% of the supernovae are thermonuclear (SNIa) and are homogeneously distributed in the Galactic disk. The Galaxy is modeled as a thin disk with a radius of 12~kpc and a height of 35~pc set from the observed scale heights of very young open clusters \citep{Hao2021a} and of O-B$_5$ stars \citep{Maiz-Apellaniz2001a}. Following \citet{Vallee2022a}, the spiral arms are described as four logarithmic spirals equally spaced in azimuth by 90$^\circ$. The pitch angle is set to 13.4$^\circ$. The starting point of the Sagittarius spiral arm is set at an angle of 50$^\circ$ and at a galactocentric distance of 2~kpc. The distribution of core-collapse supernovae in each spiral arm is finally assumed to follow a Gaussian distribution with a dispersion of 500~pc \citep{Pohl1998a}.

This simple model is used to generate 3D random distributions of SNRs in the Milky Way with random ages homogeneously drawn between 0 and $t_{\rm fade}$. An example of such a distribution is shown in Fig.~\ref{Fig-SNR-dist} (panels a and b) which displays the positions, the size, and the shock velocity of all blast waves with terminal velocities between 30 and 200~\kms\ in projection onto the Galactic disk. As shown in Fig.~\ref{Fig-SNR-dist} (panel c), any line of sight observed by \citetalias{Jenkins2011} (red lines in Fig.~\ref{Fig-SNR-dist}) intercepts either $N=0$, 1, or 2 surfaces of a given SNR, depending on the 3D position and the radius of the SNR and the 3D position of the background star. The dissipation rate induced by shocks with velocities between 30 and 200~\kms\ along this line of sight therefore simply writes
\begin{equation}
\varepsilon_{\rm diss} = \sum_{j = 1}^{\mathcal{N}_{\rm SNR}} \frac{1}{2}\,\mu\,\densini V_{B,j}^3 \,\frac{N_j}{l_{\rm los} \varphi}\, \frac{1}{{\rm cos} ( i_j )},
\end{equation}
where $i$ is the inclination angle of an SNR surface compared to the plane perpendicular to the line of sight (see Fig.~\ref{Fig-SNR-dist} panel c) and where the sum is performed over all SNRs with terminal velocities between 30 and 200~\kms.

\begin{figure}[!h]
\begin{center}
\includegraphics[width=9.0cm,trim = 1.5cm 1cm 0.5cm 2.0cm, clip,angle=0]{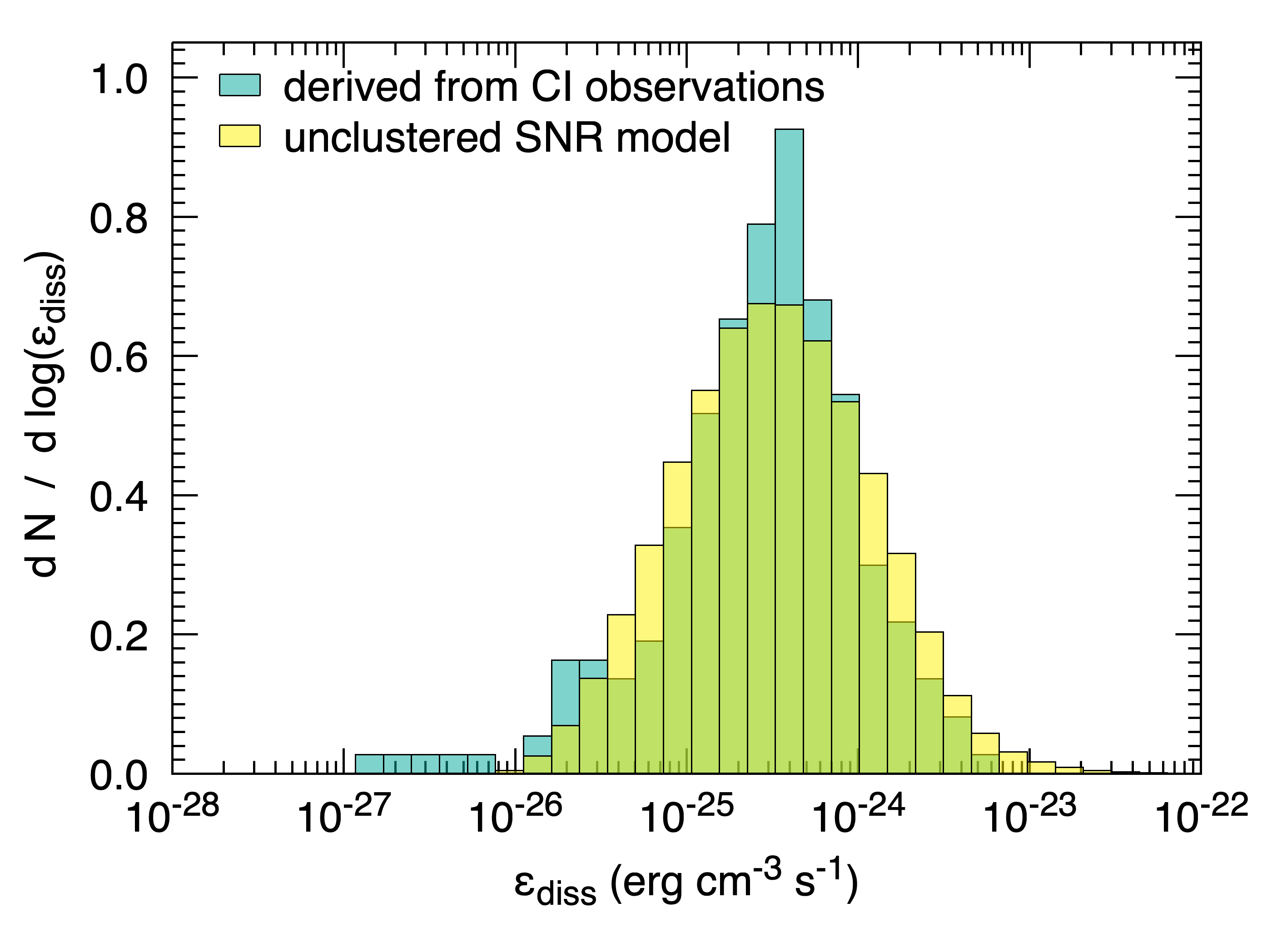}
\includegraphics[width=9.0cm,trim = 1.5cm 1cm 0.5cm 2.0cm, clip,angle=0]{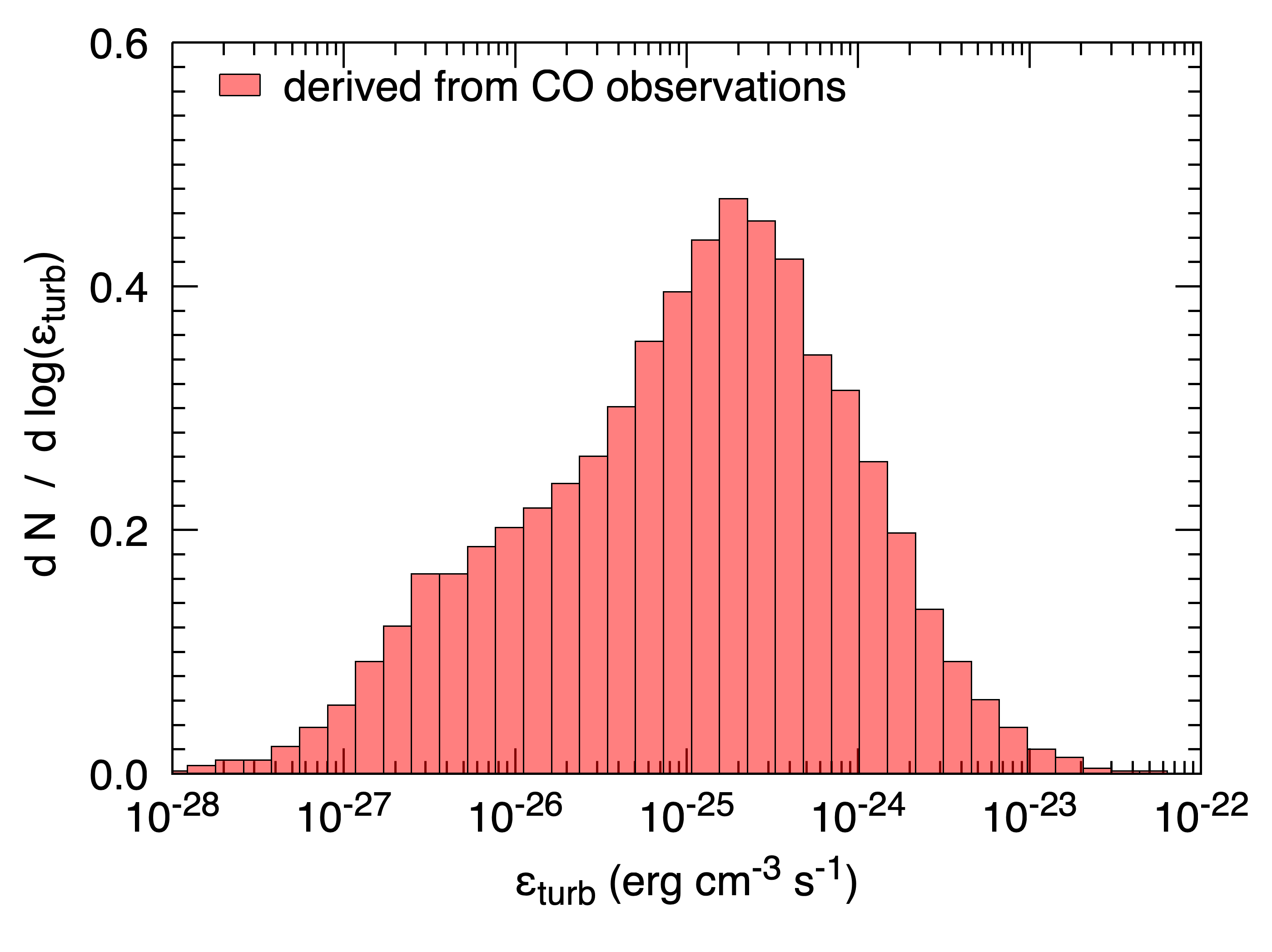}
\caption{{\it Top panel:} probability distribution function of the dissipation rate along the lines of sight observed by \citetalias{Jenkins2011} averaged over 500 realization of a simple and unclustered distribution of SNRs in the Galaxy (see Fig. \ref{Fig-SNR-dist}) compared to that derived from the production and the excitation of neutral carbon in the standard model of interstellar shocks (see Fig. \ref{Fig-eps}). {\it Bottom panel:} probability distribution function of the kinetic energy transfer rate derived from the observations of CO in molecular clouds located in the local ISM \citep{Hennebelle2012}. All distributions shown in the top and bottom panels are normalized so that their integral is equal to one.}
\label{Fig-SNR-eps}
\end{center}
\end{figure}

Fig.~\ref{Fig-SNR-eps} (top panel) displays the distribution of dissipation rates expected from SNR surfaces along the lines of sight observed by \citetalias{Jenkins2011} averaged over 500 realizations of the model described above. Interestingly, we find that shocks driven by SNRs are an unavoidable scenario. Indeed, as suggested by the projection view displayed in Fig.~\ref{Fig-SNR-dist} and as shown in Fig.~\ref{Fig-SNR-eps}, the lines of sight observed by \citetalias{Jenkins2011} are bound to intercept SNR surfaces, thus interstellar environments at high thermal pressure. The dissipation rate induced by supernova remnants is found to follow a log-normal distribution with a median value of $\sim$~$3 \times 10^{-25}$~\eccs\ and a dispersion of $\sim 0.6$ dex (corresponding to a factor of $\sim 4$). It is remarkable to see that this distribution, derived from a simplistic statistical model of SNRs, covers the same orders of magnitude than the distribution of dissipation rates required to explain the observations of the UV lines of CI (see Sect.~\ref{Sect-grid}). These results imply that shocks propagating in the WNM are not only a possible but also a probable scenario to account for the presence of gas at high thermal pressure detected in the local ISM. It also suggests that these shocks propagating at intermediate and high velocities are probably mainly driven by SNRs.

The fact that the two distributions shown in the top panel of Fig.~\ref{Fig-SNR-eps} are remarkably similar should not be overinterpreted. On the one side, it should be kept in mind that the statistical model of SNRs results from multiple simplifications. It follows that the distribution of dissipation rate derived from this model is sensitive to several parameters including the density of the ambient medium, $\densini$, the rate of supernova explosions in the Galaxy, and the geometrical parameters that describe the shape and extent of the spiral arms and of the Galactic disk. Moreover, this model builds an homogeneous distribution of core collapse supernovae along the spiral arms while supernovae are expected to explode in clusters. Since supernovae are probably spatially correlated with OB star clusters and since OB stars are the background sources used by \citetalias{Jenkins2011}, such simplification necessarily impacts the dissipation rate induced by SNR along the lines of sight observed by \citetalias{Jenkins2011}. On the other side, the decomposition procedure applied in Sect.~\ref{Sect-grid} to derive the dissipation rates from the observation of neutral carbon has several limitations. Indeed, even if the model parameters have a weak impact on the distribution of dissipation rates deduced from CI (see Sect.~\ref{Sect-grid}), adopting a distribution of shock velocities, preshock densities, and transverse magnetic field could broaden the distribution of dissipation rates up to a factor of 4 (see Fig.~\ref{Fig-eps-grd-med}) and induce a shift of the median value. Moreover, the small curvature of the line defined by the $f_1$ and $f_2$ combinations at small thermal pressure and for low velocity shocks (see Figs.~\ref{Fig-Jenkins} and \ref{Fig-f1f2-shock}) implies that the decomposition procedure is not sensitive enough to accurately measure events at low dissipation rate. This implies that the left part of the turquoise distribution shown in Fig.~\ref{Fig-SNR-eps} (for $\varepsilon_{\rm diss} \lesssim  10^{-25}$~\eccs) is less reliable than the right part.

\subsection{Connection with the turbulent cascade}

Supernovae explosions are often considered to be the main driving source of interstellar turbulence in the Milky Way. It is therefore intersting to discuss the results obtained in the previous sections in light of the properties of interstellar turbulence derived in the local ISM. One of the main properties of turbulence is the kinetic energy transfer rate of the turbulent cascade which measures the amount of energy per unit time and volume injected at large scale in turbulent motions and transferred to smaller and smaller scales where it is eventually dissipated \citep{Hennebelle2012,Miville-Deschenes2017a}. The distribution of the kinetic energy transfer rate deduced from the velocity dispersion of molecular clouds in the solar neighborhood by \citet{Hennebelle2012} is shown on the bottom panel of Fig.~\ref{Fig-SNR-eps}. In the local ISM, this transfer rate is found to vary over four orders of magnitude, with a median value $\sim 10^{-25}$~\eccs\ and a dispersion $\sim 0.8$~dex (corresponding to a factor of 6). Unexpectedly, these values are found to be similar to the median and dispersion values of the dissipation rate induced by SNRs. 

It is important to realize that the top and bottom panels of Fig.~\ref{Fig-SNR-eps} do not display the same quantities: the top panel shows the amount of energy allegedly dissipated in shocks driven by SNRs, while the bottom panel displays the amount of energy which is not dissipated in the original shocks and is allegedly injected in interstellar turbulence. The fact that these two distributions are similar may reveal a fundamental property of the mechanism by which the mechanical energy is injected in the interstellar matter. Indeed, it suggests that the amount of energy dissipated in SNRs and the amount of energy used to feed interstellar turbulence are comparable. It has been proposed in the past that turbulence might be generated in the wake of shocks through the production of baroclynic vorticity \citep{Vazquez-Semadeni1996a, Elmegreen2004} and various thermodynamical instabilities. Studying the detailed connection between the distributions shown in the top and bottom panels of Fig.~\ref{Fig-SNR-eps} requires dedicated 3D numerical simulations of supernovae blast waves. This is out of the scope of the present paper.

\section{Discussion} \label{Sect-disc2}

\subsection{Radial velocities}

This paper demonstrates that shocks propagating in the WNM are a viable scenario to explain the presence of gas at high thermal pressure detected by \citetalias{Jenkins2011}. To promote this idea, we showed that shocks efficiently produce neutral carbon with column densities, excitation conditions, and line profiles in agreement with the observations and that the energy required to power these shocks corresponds to that expected from Galactic SNRs. It is important to stress, however, that this paper does not take full advantage of all the information contained in the observations.

One additional piece of information is the distribution of radial velocities of the gas at high thermal pressure and the relation between these velocities and the ($f_1$,$f_2$) ratios. Table~\ref{Tab-obs-prof} shows that the 20 components identified as gas at high thermal pressure display a slight asymmetry of radial velocities. Indeed, eight of these 20 components have positive radial velocities up to $28$~\kms, eleven have negative radial velocities down to $-36$~\kms, and one is detected at $-83$~\kms. Moreover the positive velocity components display, on average, lower ($f_1$,$f_2$) combinations compared to the negative velocity components. This last result was already shown by \citetalias{Jenkins2011} who found that environments with the highest thermal pressure, hence highest $f_2$ ratio, are systematically associated to velocity components at least 5~\kms\ below the minimum value permitted by the differential Galactic rotation.

At first sight, this slight asymmetry would seem to contradict the simple scenario depicted in Fig.~\ref{Fig-SNR-dist} of shocks propagating in random directions and expected to have roughly equal proportions of negative and positive velocities, and to favor a scenario where part of the high pressure gas is produced through the interaction of the target stars with their immediate environments. It should be kept in mind, however, that Fig.~\ref{Fig-SNR-dist} displays an idealized picture in which SNRs are distributed homogeneously along the spiral arms and the Galactic disk. Since supernovae are spatially correlated with OB star clusters, the OB stars used as background sources by \citetalias{Jenkins2011} may preferentially reside inside one or several SNRs rather than beyond. This effect would favor the occurrence of negative velocities corresponding to the expansion of the foreground shell, and explain the fact that the gas at highest thermal pressure is preferentially associated to negative velocity components. 

Interestingly, Table~\ref{Tab-obs-prof} shows that the gas at high thermal pressure has moderate radial velocities compared to the shock velocity explored in this work. This is in line with previous observations of CI toward stars behind known SNRs which show that environments at high pressure are not always associated to significantly large displaced velocities (e.g. \citealt{Jenkins1981a, Jenkins1984a, Raymond1991a, Wallerstein1995a, Nichols2004a}). This feature could, however, be coherent with shocks driven by SNRs. Indeed, in the framework of expanding SNRs, lines of sight are more likely to intercept the borders of the expanding bubble, hence to have significant transverse motions, rather than the center of the SNR where the gas has a prominent radial motion. 

A complete modeling of the radial velocities and the asymmetry described above would require to take into account the dynamics of the Galaxy, including the rotation curve, the velocity dispersion, and the vertical motions, and the clustering of supernovae compared to OB star associations. 
This is out of the scope of the present paper.

\subsection{Limitations and alternative contributions}

The distributions depicted in Fig.~\ref{Fig-SNR-eps} suggest that shocks driven by SNRs are not only likely but also unavoidable contributors to the high-pressure gas observed in CI. However, this conclusion is moderated by the simplicity of the SNR model presented in Sect.~\ref{Sect-disc}, in particular the fact that the ambient medium in which SNRs expand is considered as homogeneous. Indeed, the interstellar medium is highly inhomogeneous and contains CNM and dense clouds that cover a great variety of spatial scales and densities. As shown in hydrodynamic, magnetohydrodynmic, adiabatic, and non-adiabatic numerical simulations, the presence of these clouds profoundly modifies the dynamics of the flow (e.g. \citealt{Pittard2010a} and references therein). The interaction of a shock with an interstellar cloud generates a forward and a reverse shocks in the cloud, a bow shock (or bow wave) upstream of the cloud structure, and reflected shocks and vorticity sheets downstream (e.g. \citealt{Poludnenko2002a,Yirak2010a}). The cloud itself is subject to Rayleigh-Taylor, Kelvin-Helmholtz, and thermal instabilities which lead to its disruption and fragmentation in the postshock turbulent flow. It follows that the expansion of an SNR in a clumpy medium could be highly non spherical (see Figs.~2~and~3 of \citealt{Korolev2015a}) and that the initial blast energy not only drives shocks in the dense clouds but is also partly converted into turbulent motions.

On the one side, the fact that part of the SNR energy is used to feed complex dynamical motions implies that the distribution of dissipation rates deduced from our simple SNR model (see Fig.~\ref{Fig-SNR-eps}) is an overestimation of the actual dissipation rate of shocks propagating in the WNM. On the other side, the propagation of low-velocity shocks in CNM structures induces an increase of the cloud thermal pressure on timescales smaller than its disruption timescale and could therefore contribute to the presence of high-pressure gas observed in CI. A complete modeling of all these aspects is of fundamental importance to understand the feedback of supernovae on the thermodynamical evolution of the multiphase ISM and will be carried in future studies.

\section{Conclusions} \label{Sect-conc}

Observations of UV absorption lines of neutral carbon in the local interstellar medium performed by \citetalias{Jenkins2011} reveals that a significant fraction of the mass of the ISM lies at thermal pressures one to three orders of magnitude above the thermal pressure of the bulk of the gas. In this paper, we propose that this enigmatic component originates from shocks propagating in the Warm Neutral Medium. To explore this idea, we perform the first detailed and quantitative study of the production and excitation of neutral carbon in atomic shocks using the latest version of the Paris-Durham shock code presented in \citetalias{Godard2024a}. The model is analyzed over a wide range of parameters that covers the typical conditions of the WNM in the local interstellar medium. We propose that these shocks are mainly driven by supernova remnants and discuss the validity of this scenario based on the expected distribution of the dissipation rate of mechanical energy. This analysis highlights the following results.
\begin{enumerate}
\item Shocks propagating in the WNM naturally generate high pressure environments. The compression of the gas favors the production of neutral carbon through the increase of the radiative and dielectronic recombination of C$^+$. Because the associated chemical timescales are much shorter than the cooling timescale, the production of neutral carbon occurs before the thermal pressure decreases.
\item The shocks explored in this work individually produce a column density of neutral carbon between $10^{12}$ and $10^{14}$~cm$^{-2}$, distributed over a Gaussian line profile with a FWHM that ranges between $1$ and $6$~\kms. These results, which  hold as long as the transverse magnetic field $B_0 \leqslant 3$~$\mu$G, are found to be in excellent agreement with the column densities and linewidths observed in velocity components solely associated to gas at high thermal pressure.
\item Although the observed line profiles suggest that individual lines of sight intercept only one or a few shocks, the current set of observations is found to be insufficient to identify their exact physical properties which include their velocities, the preshock densities, the strengths of the transverse magnetic field, and their inclination angles.
\item Each line of sight is decomposed into a high pressure and a low pressure components to compute the number of shocks required to reproduce the observations and their dissipation rates. The column densities of neutral carbon observed at high thermal pressure require a dissipation rate of mechanical energy of $\sim 3 \times 10^{-25}$~\eccs\ with a dispersion of about a factor of three. Surprisingly, this distribution of the dissipation rate weakly depends on the model parameters as long as $V_S \geqslant 30$~\kms, $\densini \geqslant 0.3$~\cc, and $B_0 \leqslant 3$~$\mu$G. This result implies that the observations of the UV absorption lines of neutral carbon provide a direct measurement of the dissipation rate induced by shocks, regardless of the properties of these shocks and their distributions along the lines of sight.
\item These shocks are likely driven by supernovae explosions and the expansion of the associated supernova remnants in the diffuse ionized and neutral phases of the Galaxy. Indeed, a simplistic statistical distribution of SNRs located in the spiral arms and in a thin Galactic disk shows that the lines of sight observed by \citetalias{Jenkins2011} inevitably crosses SNR surfaces. This model leads to an expected distribution of the dissipation rate with a median value of $\sim 3 \times 10^{-25}$~\eccs\ and a dispersion of about a factor of four, in remarkable agreement with the distribution derived from the UV lines of CI.
\item Interestingly, the distribution of the dissipation rate induced by SNRs expanding in the WNM is found to be similar to the distribution of the kinetic energy transfer rate measured in molecular clouds located in the local ISM. This result implies that the amount of mechanical energy dissipated by shocks in SNRs and the amount of energy injected in the wake of shocks in interstellar turbulence are somehow comparable.
\end{enumerate}

This work uncovers a potential direct tracer of shocks driven by supernova remnants in the nearby ISM. It also presents a methodology to estimate the rate of dissipation of kinetic energy in events that inject mechanical energy in the interstellar medium, feed the interstellar turbulence, and induce phase transition between the WNM and the CNM. The fact that shocks propagating in the WNM have a distinctive imprint on the excitation properties of neutral carbon suggest that these shocks may produce additional observational tracers. Those could be used to identify the physical properties of the shocks or their distributions and, more generally, to study the injection of mechanical energy and the phase transition process in Galactic and extragalactic environments. The search for additional observational tracers of shocks propagating in the WNM will be carried in forthcoming papers.

\begin{acknowledgements}

We are very grateful to the referee, Edward Jenkins, for his thorough reading of the manuscript and his valuable comments. The research leading to these results has received fundings from the European Research  Council, under the European Community's Seventh framework Programme, through the Advanced Grant MIST (FP7/2017-2022, No 742719). The grid of models used in this work has been run on the computing cluster Totoro of the ERC MIST, administered by MesoPSL. The models were developed using the atomic data currently available on the CHIANTI database. CHIANTI is a collaborative project involving George Mason University, the University of Michigan (USA), University of Cambridge (UK) and NASA Goddard Space Flight Center (USA). We would also like to acknowledge the support from the Programme National ``Physique et Chimie du Milieu Interstellaire'' (PCMI) of CNRS/INSU with INC/INP co-funded by CEA and CNES.

\end{acknowledgements}

\bibliographystyle{aa} 
\bibliography{mybib}

\end{document}